\documentclass[11pt]{article}
\pdfoutput=1

\usepackage{jheppub} 

\usepackage{slashed,bm}
\usepackage{array,multirow,amsmath,latexsym,bm,graphicx}
\allowdisplaybreaks 
\usepackage[normalem]{ulem}
\usepackage[T1]{fontenc}
\usepackage{multicol}

\usepackage{mathrsfs}

\def \beq{\begin{equation}}
\def \eeq{\end{equation}}
\def \beqa{\begin{eqnarray}}
\def \eeqa{\end{eqnarray}}

\def\msbar{$\overline{\hbox{MS}}$}

\def\sc{{\overline s}}

\def\Ct{{\tilde{C}}}

\def\bea{\begin{eqnarray}}
\def\eea{\end{eqnarray}}

\subheader{\small  IPPP/19/26 \\ Nikhef 2019-002}

\title{NLO corrections to $h\to b\bar b$ decay in SMEFT}

\author[a]{Jonathan~M.~Cullen,}
\author[a]{Benjamin~D.~Pecjak,}
\author[b,c]{and Darren~J.~Scott}

\affiliation[a]{Institute for Particle Physics Phenomenology, 
Durham University,  Durham DH1 3LE, UK}
\affiliation[b]{Institute for Theoretical Physics, University of Amsterdam, Science Park 904, 1098 XH Amsterdam, The Netherlands}
\affiliation[c]{Nikhef, Theory Group, Science Park 105, 1098 XG, Amsterdam, The Netherlands}

\abstract{We calculate the full set of next-to-leading order (NLO)
  corrections to $h\to b\bar{b}$ decay in the dimension-6 Standard
  Model Effective Field Theory (SMEFT).  Our calculation forms the
  basis for precision studies of this decay mode in effective field
  theory, providing analytic and numerical results for contributions
  of the 45 dimension-6 operators appearing at NLO.  On the technical
  side, we discuss several complications in NLO SMEFT computations
  which have not yet been addressed in the literature. These include
  subtleties in Higgs-$Z$ mixing,
  electric charge renormalization, and especially the treatment of
  tadpoles in SMEFT.  In particular, we highlight the role of
  decoupling relations in eliminating potentially large tadpole
  corrections to the decay rate in hybrid renormalization schemes
  which employ the \msbar~scheme for some Standard Model parameters
  (such as the $b$-quark mass and electric charge) and the on-shell
  scheme for others. }

\emailAdd{jonathan.m.cullen@durham.ac.uk}
\emailAdd{ben.pecjak@durham.ac.uk}
\emailAdd{d.j.scott@uva.nl}

\keywords{Higgs physics, Effective Field Theory}

\begin{document} 
\maketitle
\flushbottom

\section{Introduction}
\label{sec:intro}

While the discovery of the Higgs boson has been a triumph for the Standard Model (SM) of particle physics~\cite{Aad:2012tfa,Chatrchyan:2012xdj,Aad:2015zhl}, the consistency of its properties, as currently measured, with those predicted by the SM (see the experimental analyses in~\cite{Aad:2013xqa,Aad:2015mxa,Khachatryan:2016vau,Sirunyan:2017tqd,Sirunyan:2018koj} for example) has left few hints of new physics. 
An important property of the Higgs boson is its decay rate into $b$-quarks.
Despite being the largest branching fraction of the Higgs, the process $h\to b\bar{b}$ has only recently been observed by the ATLAS and CMS collaborations~\cite{Aaboud:2018zhk,Sirunyan:2018kst}. Considering the relative infancy of the Higgs measurements so far in the LHC program, as well as the prospect of future $e^+ e^-$ colliders for such studies \cite{Baer:2013cma,Peskin:2012we}, the possibility of uncovering new physics in the Higgs sector remains open. As such, the need for accurate theoretical predictions in order to correctly identify and parametrize any new physics which could be observed is paramount. 

In the absence of the direct discovery of a new particle, one possible
avenue along which to search for new physics is through the use of the
Standard Model Effective Field Theory (SMEFT). In this approach the SM
Lagrangian is supplemented with operators of mass dimension greater
than four, each with its own Wilson coefficient.  Provided the new
physics is associated with a scale $\Lambda_{\text{NP}}$ which is 
much greater than the electroweak symmetry breaking (EWSB) scale and
decouples~\cite{Appelquist:1974tg}, then its effect on processes at
low energy is captured through non-zero values of these Wilson
coefficients. This allows for a model independent approach in
attempts to identify new physics: one calculates cross sections and
decay rates within SMEFT and then fits the Wilson coefficients to data
in order to extract limits or signals of new physics.

The SMEFT operators which can be written down at a given mass
dimension are constructed out of SM fields and respect the usual SM
gauge and Lorentz symmetries. A minimal basis of operators (though not
unique) can be constructed by using the SM equations of
motion~\cite{Arzt:1993gz} and techniques to quantify the minimal
number of operators and their field content which appear at each mass
dimension have already been
developed~\cite{Henning:2015daa,Henning:2015alf,Henning:2017fpj}.  At
dimension-5 there exists only a single, lepton number violating
operator, whose Wilson coefficient is heavily suppressed.  On the other hand, at dimension-6 there are 59 independent operators for one generation of
fermions excluding baryon number violating
operators~\cite{Buchmuller:1985jz,Grzadkowski:2010es}, giving a wide
space in which to explore possible consequences for phenomenology.

Recently the inclusion of dimension-6 operators in NLO perturbative
calculations has emerged. Some general features of these calculations
have been described in
e.g.~\cite{Passarino:2012cb,Ghezzi:2015vva,deFlorian:2016spz,Passarino:2016pzb}, and the
full $59\times 59$ anomalous dimension matrix for the Wilson
coefficients needed to perform a leading-logarithmic calculation has
been calculated
in~\cite{Jenkins:2013zja,Jenkins:2013wua,Alonso:2013hga}.  At the
moment, however, there is no automated tool to produce general NLO SMEFT
predictions so these calculations are performed on a
process-by-process basis. Because of the increased complexity of these
calculations, results are available only for a handful of processes,
and often contain a limited number of operators or are restricted
to a particular set of corrections.
There are many NLO SMEFT calculations which involve a subset of 
operators~\cite{Mebane:2013zga,Chen:2013kfa,Dawson:2018jlg,Hartmann:2016pil,Vryonidou:2018eyv,Baglio:2017bfe,Baglio:2018bkm,Dawson:2018dxp},
or are restricted to QCD corrections
only~\cite{Degrande:2016dqg,Zhang:2014rja,Zhang:2013xya,Zhang:2016omx,Degrande:2018fog,Franzosi:2015osa,Bylund:2016phk,Maltoni:2016yxb,Deutschmann:2017qum,Grober:2015cwa,Neumann:2019kvk}. A
calculation of Higgs pair production at NNLO in QCD involving
dimension-6 operators which contain the Higgs field has also been
performed~\cite{deFlorian:2017qfk}. 
A small set of processes has been computed at NLO including all relevant operators in both the tree and loop level diagrams. These include lepton decay~\cite{Crivellin:2013hpa,Pruna:2014asa} and Higgs decay into vector bosons~\cite{Dawson:2018pyl,Hartmann:2015aia,Dedes:2018seb,Hartmann:2015oia,Dawson:2018liq,Dedes:2019bew}.

In this paper we obtain the full set of NLO corrections from
dimension-6 operators to the decay rate $h \rightarrow b \overline{b}$
within SMEFT.  This builds upon our previous NLO SMEFT calculations of
weak corrections in the large-$m_t$ limit or those related to four-fermion
operators \cite{Gauld:2015lmb}, and QCD
corrections~\cite{Gauld:2016kuu}. On the practical side, our
calculation forms the basis for a precision analysis of Higgs decay
into $b$-quarks within SMEFT. However, even apart from that,
calculating the full set of NLO corrections reveals features of SMEFT
beyond tree level which have not been fully addressed in the
literature. For instance, one encounters technical subtleties in the
renormalization procedure concerning electric charge renormalization
and Higgs-$Z$ and Higgs-neutral Goldstone mixing.  Moreover, when
combining electroweak and QCD corrections it is natural to introduce
hybrid renormalization schemes where some parameters are defined in
the \msbar~scheme and some in the on-shell scheme. In that case one
must pay careful attention to tadpole contributions, not only
including them in the renormalization procedure in order to obtain
gauge-independent results, but also finding a renormalization scheme
where enhanced electroweak corrections related to them are absent. In
this work we address tadpole renormalization using the ``FJ tadpole
scheme''~\cite{Fleischer:1980ub}, which is especially convenient when
performing loop calculations with automated tools, and advocate the
use of decoupling relations in building a renormalization scheme which
allows us to combine QCD and electroweak corrections in an optimal
way.

The organization of this paper is as follows. After giving an outline
of the NLO calculation as a whole in section~\ref{sec:outline}, we
describe in detail the renormalization procedure in
section~\ref{sec:Renormalization}, including our treatment of
tadpoles. We discuss sources of enhanced NLO contributions to the
decay rate in section~\ref{sec:DecouplingLargeCors}, and explain how a
hybrid renormalization scheme based on decoupling relations for the
\msbar~definition of the $b$-quark mass and electric charge is useful
when combining QCD and electroweak corrections. In section
\ref{sec:Results} we present numerical results and examine
uncertainties related to scale choices, and then conclude in
section~\ref{sec:conclusions}. We provide some details on the rotation
of the SMEFT Lagrangian to the mass basis relevant for our NLO
calculation in appendix~\ref{sec:MassBasis}, including a novel treatment of 
gauge fixing in SMEFT, and give selected analytic results for the decay rate in appendix~\ref{sec:analytic}. 
While the full analytic results are too long to print, we give them in
electronic form in the arXiv submission of this article.

\section{Outline of the calculation}
\label{sec:outline}

The dimension-6 SMEFT Lagrangian may be written as
\begin{equation}
\mathcal{L}=\mathcal{L}^{(4)}+\mathcal{L}^{(6)}; \qquad \mathcal{L}^{(6)}=\sum_i C_i(\mu) Q_i (\mu) \, ,
\end{equation}
where $\mathcal{L}^{(4)}$ denotes the SM Lagrangian, and 
$\mathcal{L}^{(6)}$ depends on the dimension-6 operators $Q_i$. We adopt the ``Warsaw basis" \cite{Grzadkowski:2010es} for these operators, which are listed in table~\ref{op59}, and the naming convention of the Wilson coefficients $C_i$  follows that of the corresponding operators. We define the Wilson coefficients such that they inherently carry two inverse powers of the new physics scale, 
$\Lambda_{\rm NP}$. 

In this paper we study the decay rate for $h\to b\bar{b}$ to NLO in SMEFT.
We can write the perturbative expansion of the decay rate up to NLO in the form
\begin{align}
\label{eq:PertExpansion}
\Gamma(h\to b\bar{b})\equiv \Gamma & = \Gamma^{(0)}+\Gamma^{(1)}  \, , 
\end{align}
where the superscripts (0) and (1) refer to the  LO and NLO contribution in perturbation theory respectively.  Each of these can be split up into SM (dimension-4) and dimension-6 contributions
with the notation
\begin{align}
\label{eq:PertExpansionSMEFT}
\Gamma^{(0)}  & = \Gamma^{(4,0)} +\Gamma^{(6,0)}  \, ,  \nonumber \\
\Gamma^{(1)} & =  \Gamma^{(4,1)} +   \Gamma^{(6,1)} \,.
\end{align}
The double superscripts $(i,j)$ refer to the dimension-$i$ 
contribution at $j$-th order in perturbation theory.  In this counting each term in $\Gamma^{(6,j)}$ contains 
exactly one Wilson coefficient of a dimension-6 operator. In other words, we allow at most one insertion of a dimension-6 operator in a given Feynman diagram and keep the interference
term of the dimension-6 amplitude with the SM, but drop the square of dimension-6 amplitude,
which is formally a dimension-8 effect at the level of the decay rate.

It is useful to divide the NLO correction from dimension-6 operators into three pieces according to 
\begin{align}
\label{eq:PertExpansion2}
\Gamma^{(6,1)}& = \Gamma^{(6,1)}_{g,\gamma}+\Gamma_t^{(6,1)} + \Gamma_{\rm rem}^{(6,1)}  \, ,
\end{align}
and analogously for the SM result $\Gamma^{(4,1)}$, which was
calculated in \cite{Kniehl:1991ze}.  The definition of the three
pieces, and the extent to which the dimension-6 corrections have been
calculated in the literature, is as follows.  First,
$\Gamma_{g,\gamma}$ contains all virtual and real emissions involving
gluons and photons.  The QCD portion of this object was calculated in
\cite{Gauld:2016kuu}.  Second, $\Gamma_t$ contains virtual weak
corrections in the large-$m_t$ limit.  These were calculated in the
on-shell renormalization scheme in \cite{Gauld:2015lmb}, where they
scale as $\alpha m_t^2/M_W^2$. Finally, the object $\Gamma_{\rm rem}$
contains the remaining virtual electroweak corrections.  The only
results available for these remaining contributions are those from
four-fermion operators obtained in \cite{Gauld:2015lmb}.

The main goal of the present work is to obtain the full NLO correction
in SMEFT.  To do this, we must calculate the UV-renormalized virtual
corrections to the LO decay rate, and add them together with real
emission corrections containing a photon or gluon.  We then evaluate
to NLO the formula
\begin{equation}
\label{eq:RealPlusVirtual}
\Gamma = \int \frac{d \phi_2}{2 m_H} | \mathcal{M}_{h \rightarrow b \bar{b}} |^2 +\int \frac{d \phi_3}{2 m_H}  |\mathcal{M}_{h \rightarrow b \bar{b}(g,\gamma)} |^2 \, ,
\end{equation}
where $d\phi_i$ is the $i$-body differential Lorentz invariant
phase-space measure.  The 2- and 3-body terms involving emissions
of gluons or photons contribute to $\Gamma_{g,\gamma}$.  These contain
IR divergences, which we regularize by performing the loop
integrations and phase-space integrals in $d=4-2 \epsilon$ dimensions.
Most of the corrections involving photons can be extracted from the
QCD calculation \cite{Gauld:2016kuu}.  The exception is real and
virtual diagrams containing a $h\gamma Z$ vertex which has no analogue
in QCD.  Analytic results for $\Gamma_{g,\gamma}$ are given in
appendix~{\ref{sec:analytic}.

The most challenging part of the calculation is to obtain the
UV-renormalized 2-body matrix element $\mathcal{M}^{(1)}(h
\rightarrow b\bar{b})$, which is needed to determine $\Gamma_{\rm rem}$.  We do this by evaluating the expression
\begin{align}
\label{eq:Mren}
\mathcal{M}^{(1)}(h \rightarrow b\bar{b}) =\mathcal{M}^{(1), \text{bare}}+\mathcal{M}^\text{C.T.} \, ,
\end{align}
where the terms on the right-hand side are the bare one-loop and counterterm amplitudes, 
respectively.  The exact form of the counterterm and bare amplitude depends on the set of independent
parameters in terms of which the SMEFT Lagrangian in the mass basis is expressed, and also 
the  scheme in which these parameters are renormalized, as discussed
in more detail below. We choose the parameters to be 
\begin{equation}
\label{eq:InputPar}
\alpha_s, \space \alpha,  \space m_f, \space m_H, \space M_W, \space M_Z,  V_{ij},\,   \space C_i  \, ,
\end{equation}
where $\alpha=e^2/(4\pi)$ and $\alpha_s=g_s^2/(4\pi)$ are the electromagnetic fine-structure and strong coupling constants respectively, and $m_f$ are the 
fermion masses.  We allow for non-vanishing third-generation masses $m_b$, $m_t$, 
and $m_{\tau}$, but set first- and second-generation fermion masses to zero. We work with the numerical
approximation of a diagonal CKM matrix $V_{ij}={\rm diag(1,1,1)}$, but do not necessarily impose Minimal Flavour Violation (MFV); further details on this point can be found in appendix~\ref{sec:Yuk_app}. 

To perform the NLO calculation, we follow the procedure set out in
\cite{Gauld:2015lmb}.  We first express the SMEFT Langrangian in the
mass basis, using the parameters in (\ref{eq:InputPar}).  There are a
number of differences in this procedure compared to the SM, the most 
significant of which involve gauge fixing,
which are described in appendix~\ref{sec:MassBasis}.  We then trade the
bare input parameters for renormalized ones in order to construct an
explicit expression for the counterterm amplitude in (\ref{eq:Mren}).
Here again there are a number of subtleties compared to the SM,
especially in the structure of tadpole contributions. The full details
of the renormalization procedure are covered in
section~\ref{sec:Renormalization}.  Finally, we must identify and
evaluate the large number of one-loop Feynman diagrams which
contribute to the bare matrix elements and UV counterterms. We have
automated the procedure by implementing the SMEFT Lagrangian in the
mass basis, including ghosts, into
\texttt{FeynRules}~\cite{Alloul:2013bka}, and then using the resulting
model file to generate the diagrams with
\texttt{FeynArts}~\cite{Hahn:2000kx} and compute them with
\texttt{FormCalc}~\cite{Hahn:1998yk}. We have also made use of
\texttt{Package-X}~\cite{Patel:2015tea} when extracting analytic
expressions for loop integrals.

The NLO correction $\Gamma^{(1)}$ obtained in this way is quite lengthy.  In fact, we obtain 
contributions from 45 different dimension-6 operators when full mass dependence of third-generation 
fermions is kept.  We give the result in symbolic form in the computer files available with the electronic version of this submission.    We have performed three main checks on these results.  The first is that the UV poles in the bare and counterterm matrix elements cancel against each other, and the related fact that the decay 
rate is independent of the renormalization scale $\mu$ up to NLO.  
The second is that the IR poles appearing in the 2- and 3-body contributions to $\Gamma^{(1)}_{g,\gamma}$ cancel against each other.  Finally, we have verified the gauge independence of our 
results by performing  all calculations in both unitary and Feynman gauge.

\section{The renormalization procedure}
\label{sec:Renormalization}

In this section we lay out the renormalization procedure used in our calculation. We draw 
on the methods used in~\cite{Gauld:2015lmb} to construct the one-loop counterterm
in section~\ref{sec:CTS}, but must deal with technical complications not present in the partial 
NLO calculation in the  on-shell scheme performed there. We point out 
subtleties with charge renormalization  in section~\ref{ECR} and with Higgs-$Z$ 
mixing in section~\ref{sec:HiggsMixing}, before moving on to discuss tadpole
renormalization in section~\ref{sec:Tadpoles}.

\subsection{The one-loop counterterm}
\label{sec:CTS}

The form of the NLO counterterm follows directly from the LO  decay amplitude.  
We write the LO decay amplitude as  
\begin{align}
i{\cal M}^{(0)}(h\to b\bar b)= -i \bar u(p_b) 
\left({\cal M}_{L}^{(0)} P_{L} + {\cal M}_{L}^{(0)*} P_R\right)v(p_{\bar b}) \, ,
\end{align}
which we split up as 
\begin{align}
{\cal M}_{L}^{(0)} = {\cal M}_{L}^{(4,0)}+{\cal M}_{L}^{(6,0)} \, ,
\end{align}
where the superscripts $(4,0)$ and $(6,0)$ refer to the dimension-4 and dimension-6 contributions respectively. In order to express results in terms of our choice of input parameters (\ref{eq:InputPar}), it is convenient to introduce
\begin{align}
\label{eq:SMVar}
\hat{v}_T \equiv \frac{2 M_W \hat{s}_w}{e}, \qquad \hat{c}_w^2 \equiv \frac{M_W^2}{M_Z^2}, \qquad 
\hat{s}_w^2 \equiv 1-\hat{c}_w^2 \, .
\end{align}
These hatted quantities are defined in terms of masses and couplings as in the SM.  
After rotation to the mass basis following the steps in appendix~\ref{sec:MassBasis} one finds
\begin{align}
\label{eq:Mtree}
{\cal M}_{L}^{(4,0)} &= \frac{m_b}{\hat{v}_T} \, , \\
\label{eq:Mtree6}
{\cal M}_{L}^{(6,0)} &= 
m_b\hat{v}_T \left[ C_{H\Box}-\frac{C_{HD}}{4}\left(1- \frac{\hat{c}_w^2}{\hat{s}_w^2} \right)
+\frac{\hat{c}_w}{\hat{s}_w} C_{HWB}  
-\frac{\hat{v}_T}{m_b} \frac{C_{bH}^*}{\sqrt{2}}\right] \, .
\end{align} 
Our notation is such that $C_{fH}$ is the coefficient which contributes to 
the  $hff$ coupling after rotating to the mass basis. Its precise definition in terms of the coefficients 
multiplying the weak-basis operators in table~\ref{op59} can be found in appendix~\ref{sec:Yuk_app}.

The LO decay amplitude (as well as the NLO counterterm derived from it) depends on the
choice of input parameters. Using those given in (\ref{eq:InputPar}) requires that we eliminate 
$v_T$ according to the relation~\cite{Gauld:2015lmb}
\begin{align}
\label{eq:vevrep}
\frac{1}{v_T} = \frac{1}{\hat{v}_T}\left(1 + \hat{v}_T^2 \frac{\hat{c}_w}{\hat{s}_w}\left[C_{HWB}+\frac{\hat{c}_w}{4\hat{s}_w}C_{HD} \right]\right) \, ,
\end{align}
as has already been done in (\ref{eq:Mtree6}). In contrast, in the $G_F$-scheme, which was used when calculating the partial NLO
results of \cite{Gauld:2015lmb}, one employs
\begin{align} \label{eq:GF}
 \frac{1}{\sqrt{2}} \frac{1}{v_T^2}= G_F
-\frac{1}{\sqrt{2}}\left(C_{\substack{Hl \\ ee}}^{(3)}+C_{\substack{Hl \\ \mu\mu}}^{(3)} \right) + \frac{1}{2\sqrt{2}}
\left(C_{\substack{ll \\ \mu ee \mu}}
+ C_{\substack{ll \\ e\mu \mu e}} \right) \, .
\end{align}
We have found the choice (\ref{eq:vevrep}) to be particularly
convenient for the full NLO calculation, since it involves only
parameters which appear in the Lagrangian, and no tree-level dependence on
four-fermion operators contributing to muon decay is introduced.
Of course, it is a simple matter to convert the results obtained here  to other renormalization schemes, 
provided all finite shifts between input parameters are known
completely to NLO in SMEFT -- for instance, calculations needed to trade $v_T$ for  $G_F$ as in~(\ref{eq:GF}) 
have been obtained in~\cite{Dawson:2018pyl}.

The NLO counterterm is obtained by interpreting the  external fields and parameters
in (\ref{eq:Mtree}) and~(\ref{eq:Mtree6}) as bare  ones, which are then replaced by renormalized ones before
expanding the resulting expression to NLO in the couplings.  The bare and renormalized
fields are related through wavefunction renormalization factors according to
\begin{align}
\label{eq:wfRen}
h^{(0)} &=\sqrt{Z_h} h = \left(1+\frac12 \delta Z_h \right)h  \,, \nonumber \\
b^{(0)}_L & = \sqrt{Z^{L}_b}b_L = \left(1+\frac12 \delta Z^L_b\right)b_L \,, \nonumber \\
b^{(0)}_R &= \sqrt{Z^{R}_b} b_R = \left(1+\frac12 \delta Z^R_b\right)b_R \, ,
\end{align}
where the second equality on each line is valid to NLO. 
For the masses, electric charge, and Wilson coefficients we write
\begin{align}
\label{eq:parRen}
M^{(0)} = M + \delta M, \qquad e^{(0)} = e + \delta e, 
\qquad C_i^{(0)} =C_i + \delta C_i\, ,
\end{align}
where $M$ is a generic mass. The bare quantities in (\ref{eq:wfRen}) and (\ref{eq:parRen})  
are labeled with  a  superscript $(0)$ while the renormalized ones are not, and the counterterm
for an arbitrary quantity $X$ is denoted by $\delta X$.  These NLO counterterms are calculated in 
perturbation theory and receive both dimension-4 and dimension-6 contributions, which we 
denote by $\delta X^{(4)}$ and $\delta X^{(6)}$, respectively.  

 Inserting these expressions into 
(\ref{eq:Mtree}) and (\ref{eq:Mtree6}) and keeping only the linear terms in $\delta X$ gives an expression for the NLO
counterterm for the decay amplitude.  Writing this as 
\begin{align}
i \mathcal{M}^{\text{C.T}}(h\rightarrow b \bar{b})=-i \bar{u}(p_b) \left( \delta \mathcal{M}_L P_L + \delta \mathcal{M}_L^* P_R \right) v(p_{\bar{b}} ) \, ,
\end{align} 
the dimension-4 counterterm is 
\begin{align}
\label{eq:dim4CT}
\delta {\cal M}_L^{(4)} & = \frac{m_b}{\hat{v}_T}
\left(\frac{\delta m_b^{(4)}}{m_b}-\frac{\delta \hat{v}_T^{(4)}}{\hat{v}_T}+\frac{1}{2} \delta Z_h^{(4)} +\frac{1}{2}\delta Z_b^{(4),L} + \frac{1}{2}\delta Z_b^{(4),R*} \right)  \, ,
\end{align}
while the  dimension-6 counterterm is 
\begin{align}
\label{eq:dim6CT}
\hspace{0.6cm} \delta {\cal M}_L^{(6)} =& 
\frac{m_b}{\hat{v}_T}\left(\frac{\delta m_b^{(6)}}{m_b} 
-\frac{\delta \hat{v}_T^{(6)}}{\hat{v}_T}+\frac12 \delta Z_h^{(6)}+ \frac12 \delta Z_b^{(6),L}+ \frac{1}{2} \delta Z_b^{(6),R*}\right) 
\nonumber \\*
&+
{\cal M}_L^{(6,0)}  \left(\frac{\delta m_b^{(4)}}{m_b} 
+\frac{\delta \hat{v}_T^{(4)}}{\hat{v}_T}+\frac12 \delta Z_h^{(4)}+ \frac12 \delta Z_b^{(4),L}+ \frac{1}{2} \delta Z_b^{(4),R*}\right) \nonumber \\* 
&- \frac{\hat{v}_T^2}{\sqrt{2}} C_{bH}^*  \left(\frac{\delta \hat{v}_T^{(4)}}{\hat{v}_T}-\frac{\delta m_b^{(4)}}{m_b}  \right) 
+m_b \hat{v}_T \left[C_{HWB}+\frac{\hat{c}_w}{2\hat{s}_w}C_{HD} \right]\delta\left(\frac{\hat{c}_w}{\hat{s}_w}\right)^{(4)}
 \nonumber \\*
& +  m_b \hat{v}_T\left( \delta C_{H\Box}-\frac{\delta C_{HD}}{4}\left(1- \frac{\hat{c}_w^2}{\hat{s}_w^2} \right)
+\frac{\hat{c}_w}{\hat{s}_w} \delta C_{HWB}  
-\frac{\hat{v}_T}{m_b} \frac{\delta C_{bH}^*}{\sqrt{2}}\right)   \, ,
\end{align}
where we have defined 
\begin{equation}
\frac{\delta \hat{v}_T}{\hat{v}_T} \equiv \frac{\delta M_W}{M_W}+\frac{\delta \hat{s}_w}{\hat{s}_w}-\frac{\delta e}{e} \, .
\end{equation}
From the definitions of $\hat{c}_w$ and  $\hat{s}_w$ in (\ref{eq:SMVar}) one finds that
\begin{equation}
\frac{\delta \hat{s}_w}{\hat{s}_w} = 
-\frac{\hat{c}_w^2}{\hat{s}_w^2} 
\left(\frac{\delta M_W}{M_W}- \frac{\delta M_Z}{M_Z}\right) \, , 
\quad
\delta \left(\frac{\hat{c}_w}{\hat{s}_w} \right)^{(4)}= -\frac{1}{\hat{c}_w\hat{s}_w}\left(\frac{\delta \hat{s}_w^{(4)}}{\hat{s}_w}\right)
 \, .
\end{equation}
The NLO counterterms are computed by specifying a 
renormalization scheme and evaluating one-loop Feynman diagrams as appropriate in that scheme. 
For the Wilson coefficients, we use the \msbar~scheme, where the counterterms 
involve only UV poles in the dimensional regulator 
$\epsilon = (4-d)/2$.\footnote{In fact, counterterms in the \msbar~scheme are proportional
to  $\frac{1}{\epsilon}- \gamma_E + \ln(4\pi)$,  but since the finite 
terms cancel from renormalized amplitudes along with the UV poles we omit them for simplicity.}     
In that case, we can read off the NLO counterterms from the anomalous 
dimension calculation performed in 
\cite{Jenkins:2013zja, Jenkins:2013wua, Alonso:2013hga}.  
The counterterms take the  form 
\begin{align}
\label{eq:delC}
\delta C_i =\frac{1}{2 \epsilon} \dot{C}_i(\mu)  \,,
\end{align}
where we have introduced 
\begin{align}
\label{eq:delCdef}
\dot{C}_i(\mu) \equiv  \mu \frac{d}{d\mu}C_i(\mu) = \sum_j \gamma_{ij} C_j  \,,
\end{align}
with $\gamma_{ij}$ the anomalous dimension matrix. In general $\gamma_{ij}$ is not 
diagonal, so any Wilson coefficient counterterm is a linear combination of many other Wilson coefficients in the chosen basis.

The wavefunction, mass, and electric charge counterterms are determined by calculating 
a set of one-loop integrals in the mass basis.  The construction of these counterterms in 
SMEFT closely follows the procedure used in the Standard Model, as outlined, for instance, in 
\cite{Denner:1991kt}.  Most of the details needed for $h\to b\bar{b}$ decay in the on-shell 
scheme were given in \cite{Gauld:2015lmb}.  However, while wavefunction renormalization factors 
are always evaluated on-shell, in the present work we aim to be flexible in the treatment 
of mass and electric charge renormalization, allowing for hybrid schemes which define
some of these parameters in the on-shell scheme, and some in the \msbar~scheme. In that case
we must pay careful attention to tadpole contributions, as explained in section~\ref{sec:Tadpoles}. There are also some subtleties in electric charge renormalization and Higgs-$Z$ mixing once
dimension-6 effects are included, which we cover in sections~\ref{ECR} and~\ref{sec:HiggsMixing} below.
  
When necessary, we distinguish parameters in the on-shell scheme from those in the \msbar~scheme
through the notation
\begin{align}
X^{\rm O.S.}= X^{(0)} + \delta X^{{\rm O.S.}} \, , \nonumber  \\
\overline{X}(\mu) = X^{(0)}  + \delta \overline{X}(\mu) \, , \label{eq:MSbarX}
\end{align}
where O.S.~indicates the on-shell scheme and we have made the $\mu$ dependence in 
the \msbar~parameter $\overline{X}(\mu)$ explicit.  The counterterms in the two 
schemes have the same UV divergences, but differ in the finite parts: 
the UV-finite part is set to zero in the \msbar~scheme and determined through 
on-shell renormalization conditions in the on-shell scheme. 
We can therefore facilitate conversion between the \msbar~and on-shell schemes by writing
 \begin{align}
\label{eq:CXdef}
X =X^{(0)}\left(1+ \frac{\delta X^{\rm div.} }{X} + c_{X} \frac{\delta X^{\rm O.S., fin.} }{X}   \right) \, ,
\end{align}
where the notation splits the counterterm into UV-divergent $(X^{\rm div.})$ and UV-finite 
$(\delta X^{\rm fin.})$ pieces. Results in the on-shell scheme are picked out by setting
$c_{X}=1$,  while  $c_{X}=0$ picks out the \msbar~scheme. This notation
allows us to suppress  the extra labels in (\ref{eq:MSbarX}) and refer instead to a 
generic quantity $X$, with the understanding that the renormalization scheme can be
specified by adjusting the value of $c_{X}$ and the numerical value of $X$ appropriately.
We use this notation in section~\ref{sec:DecouplingLargeCors} and appendix~\ref{sec:analytic}.

\subsubsection{Electric charge renormalization}
\label{ECR}
\label{sec:WI}
The one-loop counterterm~(\ref{eq:dim6CT}) involves both SM and dimension-6 contributions from
electric charge renormalization. The SM calculation simplifies due to electroweak Ward identities, 
which relate the $ff\gamma$ vertex function to two-point functions through gauge invariance. 
Adapting the notation of \cite{Denner:1991kt} to our conventions, these  allow one to write
\begin{align}
 \label{eq:Charge} 
 \frac{\delta e^{(4)}}{e} = \frac{1}{2}
\frac{\partial \Sigma_T^{AA(4)}(k^2)}{\partial k^2} \bigg|_{k^2=0} -
\frac{(v_f^{(4)} - a_f^{(4)})}{Q_f}\frac{\Sigma_T^{AZ(4)}(0)}{M_Z^2} \, ,
\end{align}
where as usual the superscript $(4)$ refers to dimension-4 contributions. The object
$\Sigma_T^{AA}$ ($\Sigma_T^{AZ}$) is the transverse component of the
$\gamma\gamma$ ($\gamma Z$) two-point function.  The $\gamma Z$ two-point
function is needed for charge renormalization in the SM because the photon can mix into
a $Z$-boson through loop corrections before coupling to the fermion, and it is for the same reason that
the axial-vector ($a_f$) and vector ($v_f$) couplings of the $Z$-boson to fermions enter
the expression. In the SM $v_f^{(4)} - a^{(4)}_f=-Q_f\hat{s}_w/\hat{c}_w$, which makes explicit the important
feature that $\delta e$ is independent of the fermion $f$.  

To renormalize the $h\to b\bar{b}$ decay amplitude we also need 
the dimension-6 counterterm $\delta e^{(6)}$. We have determined this expression
by renormalizing the $ff\gamma$ vertices directly, without using the SM Ward identities. 
We find by explicit calculation that 
\begin{align}
\label{eq:ChargeSMEFT}
 \frac{\delta e^{(6)}}{e} = \frac{1}{2}
\frac{\partial \Sigma_T^{AA(6)}(k^2)}{\partial k^2} \bigg|_{k^2=0} +
\frac{1}{M_Z^2}\left(\frac{\hat{c}_w}{\hat{s}_w} \Sigma_T^{AZ(6)}(0)   -\frac{\hat{v}_T^2}{4\hat{c}_w\hat{s}_w}C_{HD} \Sigma_T^{AZ(4)}(0) \right)\, .
\end{align}
Although the counterterm can be obtained from two-point functions alone,
one can verify that the term multiplying $\Sigma_T^{AZ}$ differs from the
form $(v_f-a_f)/Q_f$ through terms involving the class-7 operators $Q_{Hf}$.  
In fact, since $v_f^{(6)}=-a_f^{(6)} = C_{Hf} \hat{v}_T^2/{4 \hat{c}_w\hat{s}_w}$ for these 
operators, a naive generalization of the SM result (\ref{eq:Charge}) would lead to the contradictory
result that electric charge renormalization depends on the fermion charge $Q_f$.  
An important check on this  expression is that the UV poles in the NLO decay amplitude cancel once it is used.

\subsubsection{Higgs-\texorpdfstring{$Z$}{Z} mixing}
\label{sec:HiggsMixing}

In general, the SMEFT Wilson coefficients contain imaginary parts even after writing the Lagrangian in the mass
basis. While these drop out of the NLO  the decay rate, they appear in the NLO decay amplitude 
and introduce complications into the renormalization procedure which are irrelevant in the SM.  
One of these is mixing of the SM Higgs field $h$  with the longitudinal component of
the $Z$-boson and the neutral Goldstone boson $\phi^0$ (in $R_\xi$ gauge) at the one-loop level. 
Since $h$  and $\phi^0$ are the real and 
imaginary parts of the neutral component of the Higgs doublet $H$ after electroweak symmetry
breaking respectively (see e.g. (\ref{eq:HiggsDoublet})), this mixing must involve a complex coupling.  
However, in the SM  neutral-current couplings are real after transformation to the mass basis,  
so there is no such mixing at NLO. In SMEFT, however, diagrams of the type 
shown in Figure~\ref{fig:HiggsMixing}  contribute to the $h \rightarrow b \bar{b}$ decay amplitude, 
where $f$ is any massive fermion.  The sum of diagrams yields a gauge-invariant result proportional
to 
\begin{align}
\eta_5 = \frac{ \sqrt{2}}{\hat{v}_T} \, {\rm Im}\left[N_c m_b C_{bH} - N_c m_t C_{tH}+m_{\tau}C_{\tau H}+\dots \right] \, ,
\end{align}
where the $\dots$ refer to contributions from second- and third-generation fermions, which take on the same structure. The loop
integrals multiplying $\eta_5$ contain UV divergences which are
exactly canceled by the piece in the Wilson coefficient counterterm
(\ref{eq:delC}) involving $\dot{C}_{bH}$, which was calculated with
the SMEFT Langrangian in the unbroken phase of the theory (i.e. when
the vacuum expectation value of the Higgs field vanishes) in
\cite{Jenkins:2013wua}.

While in the unbroken phase it is unambiguous that the $\eta_5$ term
arises from mixing of real and imaginary parts of the complex Higgs
doublet, in the broken phase the exact origin (but not the result
itself) depends on the gauge: in unitary gauge it is due entirely to
Higgs mixing with the longitudinal component of the $Z$-boson, while in
$R_\xi$ gauge it is due to the sum of graphs containing $Z$ and
neutral Goldstone bosons.\footnote{B.P.\ is grateful for a discussion
  with Aneesh Manohar which clarified this point.}

\begin{figure}[t]
\centering
\[
\begin{array}{ccc}
\includegraphics[scale=0.85]{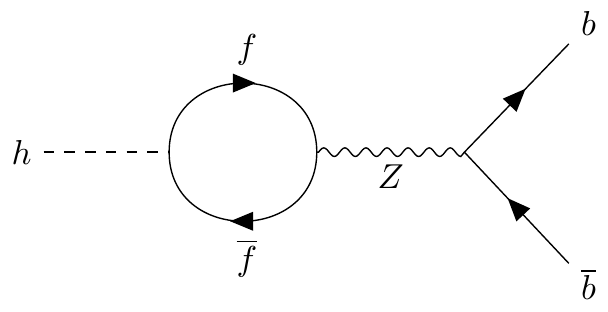} & \includegraphics[scale=0.85]{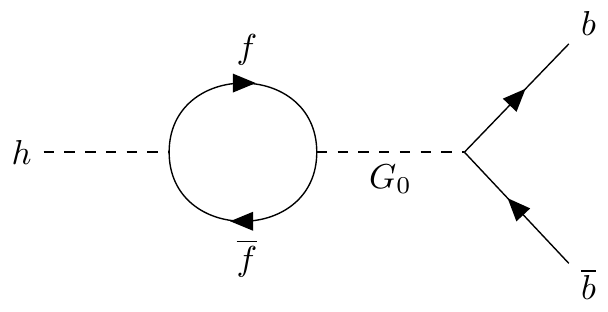} & \includegraphics[scale=0.85]{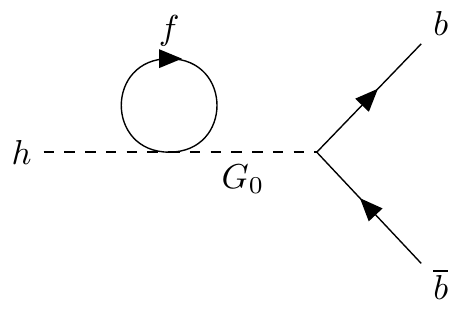} \\
(a) & (b) & (c) \\
\end{array}
\]
\caption{Diagrams contributing to the partial width of $h \rightarrow b \overline{b}$ from Higgs mixing to (a) $Z$-boson and (b,c) neutral Goldstone boson.}
\label{fig:HiggsMixing}
\end{figure}

\subsection{Tadpoles}
\label{sec:Tadpoles}

In the on-shell renormalization scheme tadpole contributions cancel between different terms in the 
renormalized amplitude.  For this reason,  no tadpoles were included in 
the partial NLO calculation in the on-shell scheme in \cite{Gauld:2015lmb}.  However, if some 
parameters are renormalized in the on-shell scheme and some in the \msbar~scheme, then tadpole
cancellations only happen at the level of UV-divergent parts of the amplitudes.  Tadpoles remain 
in the finite parts, and must be taken into account to arrive at a gauge-invariant result.  In 
fact, only upon the inclusion of tadpoles are the one-loop matrix elements (including wavefunction renormalization
factors) and also mass and parameter counterterms individually gauge invariant  \cite{Actis:2006ra}. 

There are various schemes for the treatment of tadpoles available in
the literature.  We have chosen to perform our calculations using the
so-called ``FJ tadpole scheme" \cite{Fleischer:1980ub}, an excellent
discussion of which is given in \cite{Denner:2016etu}.\footnote{As
  described in \cite{Denner:2016etu}, this scheme is closely related
  to the $\beta_t$ scheme of \cite{Actis:2006ra}.}  As explained in
that paper, a property of the FJ tadpole scheme is that it is equivalent
to a scheme where tadpoles are not renormalized.  In other words,
tadpole renormalization can be taken into account simply by including tadpole
topologies into any $n$-point amplitude entering a given calculation.
This scheme applies not only to the Standard Model, but rather to
generic theories, therefore it extends to SMEFT with no essential
complications. We find this scheme to be particularly convenient,
since it means that instead of adding explicit tadpole counterterms to
the already lengthy expression (\ref{eq:dim6CT}), we need only include
tadpole topologies into our diagrammatic calculations, which in any
case have been automated.
\begin{figure}[t]
\begin{minipage}[l][7cm][c]{7cm} 
\begin{center}
\[
\begin{array}{cc}
\includegraphics[scale=1]{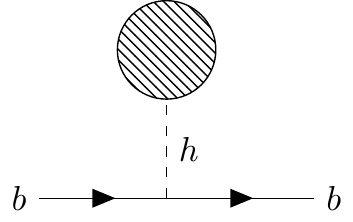} & \includegraphics[scale=1]{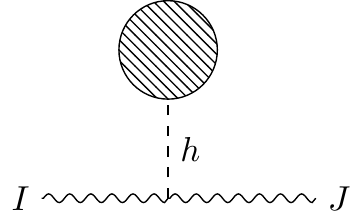} \\
(a) & (b) \\
\, & \, \\
\includegraphics[scale=1]{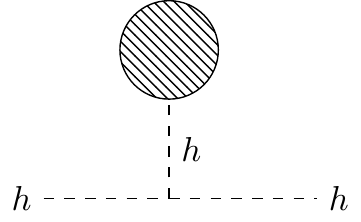} & \includegraphics[scale=1]{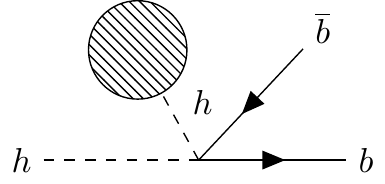} \\
(c) & (d) \\
\end{array}
\]
\end{center}
\end{minipage}
\hspace{1cm}
\begin{minipage}[r][7cm][c]{7cm} 
\begin{center}
\[
\begin{array}{cc}
\includegraphics[scale=1]{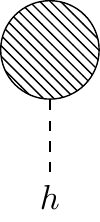} \\
(e)
\end{array}
\]
\end{center}
\end{minipage}
\caption{\label{fig:TadpoleAll}
NLO tadpole diagrams which appear in our calculation. In addition to 
contributions to two-point functions of  (a) the $b$-quark, (b)  vector bosons, where
$IJ = \gamma\gamma, \space \gamma Z, \space WW, \space ZZ$, and (c)  the 
Higgs, the contributions to the $h\to b\bar{b}$ matrix element shown in (d) appear through the 
dimension-6 operator $Q_{bH}$.
In each case the diagram factorizes into the product of the tadpole function in (e) with a  
Higgs propagator and a Higgs coupling to
the tree-level diagram. }
\end{figure}

In $h\to b\bar{b}$ decay within the SM, tadpole contributions appear in the two-point functions used
for mass and parameter renormalization through the diagrams shown in Figure~\ref{fig:TadpoleAll}(a)--(c). 
In $h\to b\bar{b}$ decay within SMEFT, tadpoles appear not only in the two-point functions, but also in the bare decay amplitude 
through the diagram shown in Figure~\ref{fig:TadpoleAll}(d).  
We can write any of these diagrams as the product of the one-point tadpole function $T$ shown in
Figure~\ref{fig:TadpoleAll}(e) with a tree level graph, provided we include the appropriate
Higgs coupling and propagator. We write the result for the  tadpole function  
\begin{align}
T=   T^{(4)}+T^{(6)} \, ,
\end{align}
where $(4)$ and $(6)$ represent the SM and dimension-6 contributions respectively. In unitary
 gauge one has
\begin{align}
T_{\rm un.}^{(4)} = \frac{1}{32 \pi^2 \hat{v}_T } \Bigg\{&
6\left(1-\frac{2\epsilon}{3}\right)\left[2 M_W^2 A_0(M_W^2) +M_Z^2 A_0(M_Z^2)\right]
+3 m_H^2 A_0(m_H^2)
\nonumber \\* & 
-8 \sum_{f} N_c^f m_{f}^2 A_0(m_{f}^2) \Bigg\} \, ,
\end{align}
while in Feynman gauge 
\begin{align}
\label{eq:TdiffSM}
T_{\rm Feyn.}^{(4)} = T_{\rm un.}^{(4)} + 
\frac{m_H^2}{32\pi^2 \hat{v}_T} \left[2 A_0(M_W^2) +A_0(M_Z^2)\right] \, , 
\end{align}
where $f$ refers to quarks ($q$) or charged leptons ($l$) with $N_c^q=3$, $N_c^l=1$, and 
\begin{align}
A_0(M^2) =M^2\left(\frac{1}{\epsilon}+ \ln\left(\frac{\mu^2}{M^2}\right)+1 \right) \,.
\end{align}
For the dimension-6 contribution in unitary gauge we find 
\begin{align}
T_{\rm un.}^{(6)} = 
\frac{ \hat{v}_T}{32 \pi^2 } &\Bigg\{\left(-6 C_H \hat{v}_T^2  +4 C_{H,{\rm kin}}\frac{m_H^2}{\hat{v}_T^2}\right)A_0(m_H^2) +(24-16\epsilon) C_{HW} M_W^2 A_0(M_W^2) 
\nonumber \\* &
+ \left(3- 2\epsilon\right)\left[ C_{HD} +4(C_{HW} \hat{c}_w^2+C_{HB} \hat{s}_w^2 +\hat{c}_w \hat{s}_w C_{HWB}) \right]M_Z^2 A_0(M_Z^2)
\nonumber \\* &
+\sum_f N_c^f 2\sqrt{2}\hat{v}_T m_f (C_{fH}+C^*_{fH})A_0(m_f^2)
\Bigg\}
\nonumber \\* &
+\left[C_{H,\rm kin}
+\hat{v}_T^2 \frac{\hat{c}_w}{\hat{s}_w}\left(C_{HWB}+\frac{\hat{c}_w}{4\hat{s}_w}C_{HD} \right)\right]
T^{(4)}_{\rm un.} \, ,
\end{align}
and in Feynman gauge
\begin{align}
T_{\rm Feyn.}^{(6)}&=T_{\rm un.}^{(6)} - \frac{m_H^2 \hat{v}_T}{16 \pi^2 }
 \left(\frac{2}{\hat{v}_T^2} C_{H,\rm kin}A_0(M_W^2)+C_{H\Box}A_0(M_Z^2)  \right)
  \nonumber \\
 &+ \left[C_{H,\rm kin}+\hat{v}_T^2 \frac{\hat{c}_w}{\hat{s}_w}\left(C_{HWB}+\frac{\hat{c}_w}{4\hat{s}_w}C_{HD} \right)\right]
(T^{(4)}_{\rm Feyn.}-T^{(4)}_{\rm un.}) \,,
\end{align}
where $C_{H,\rm kin}$ is defined in (\ref{eq:CHkin}).

An interesting feature of SMEFT is that, in contrast to the SM, tadpole diagrams contribute to 
electric charge renormalization through the $\gamma\gamma$ two-point function.  These 
contributions are proportional to the $h\gamma\gamma$ coupling in SMEFT, which is induced
by class-4  operators and involves the combination of Wilson coefficients
\begin{align}
\label{eq:cHGG}
c_{h\gamma\gamma}=  C_{HB} \hat{c}_w^2+C_{HW} \hat{s}_w^2- C_{HWB} \hat{c}_w \hat{s}_w  \,.
\end{align}
Direct calculation in unitary gauge of the piece of the electric charge counterterm as 
described in section~\ref{sec:WI}  yields the result
\begin{align}
\frac{\delta e^{{\rm cl.4},(6)}}{e} = \frac{1}{16\pi^2}\left[c_{h\gamma\gamma}A_0(m_H^2) +
4 \hat{c}_w \hat{s}_w C_{HWB}\left(4M_W^2 -3 A_0(M_W^2) \right) \right]-
2 c_{h\gamma\gamma}\frac{\hat{v}_T}{m_H^2} T^{(4)}_{\rm un.}\,,
\end{align}
where the extra superscript ``cl.4" indicates restriction to class-4 operators in table~\ref{op59}.  The term proportional to the SM tadpole function $T^{(4)}_{\rm un.}$ arises through diagrams of the type 
shown in Figure~\ref{fig:TadpoleAll}(b) with $IJ=\gamma\gamma$.  In Feynman gauge the division into 
tadpole and the remaining contributions reads instead
\begin{align}
\label{eq:deltaEC4Feyn}
\frac{\delta e^{{\rm cl.4},(6)}}{e} = \frac{1}{16\pi^2}&\left[c_{h\gamma\gamma}\left( A_0(m_H^2) + 2A_0(M_W^2)+A_0(M_Z^2)\right)
+4  \hat{c}_w \hat{s}_w  C_{HWB}\left(4M_W^2 -3 A_0(M_W^2) \right) \right]
\nonumber \\ &
-2 c_{h\gamma\gamma}\frac{\hat{v}_T}{m_H^2} T^{(4)}_{\rm Feyn.}\,,
 \end{align}
but the end result is the same due to  (\ref{eq:TdiffSM}).  

This example illustrates the general feature that parameter counterterms are gauge invariant only after 
including tadpoles. The same is true of the sum of bare matrix elements and wavefunction
renormalization factors, which is also a gauge-invariant object. The mechanism 
through which tadpoles ensure this gauge invariance is rather non-trivial.  
For instance, in contrast to the SM, tadpoles contribute directly to bare matrix elements through 
diagrams of the type shown in Figure~\ref{fig:TadpoleAll}(d).  They also contribute to wavefunction 
renormalization of the $b$-quark field.   Evaluating the tadpole contribution to the $b$-quark
self-energy shown in Figure~\ref{fig:TadpoleAll}(a) and using it to extract the wavefunction 
renormalization factor using the convention of \cite{Gauld:2015lmb}, one finds
\begin{align}
\delta Z^L_{b,{\rm tad.}} = - \frac{i \sqrt{2} \hat{v}_T^2}{m_H^2 m_b} \text{Im}(C_{bH}) T^{(4)} \, ,
\end{align}
where $T$ is the tadpole function in the chosen gauge.  While this purely imaginary contribution 
drops out of the NLO decay rate, it is needed to ensure gauge invariance of the sum of the NLO  
matrix element and the wavefunction renormalization factors, and also plays a role in the
cancellation of tadpoles in the on-shell scheme.  

These examples illustrate that while the treatment of tadpoles in SMEFT is conceptually the same
as in the SM, the exact structure of tadpoles in the diagrammatic calculations 
is more involved.  We have calculated all tadpole 
contributions to the bare matrix elements and counterterms appearing in the $h\to b\bar{b}$ decay
amplitude at NLO in unitary gauge and in Feynman gauge, and confirmed that the gauge dependence 
in the tadpole functions cancels against that in other diagrams, such that the counterterms for mass and electric charge renormalization, as well as the sum of the bare matrix element and the wavefunction renormalization factors, are 
separately gauge invariant.  

We have also confirmed that tadpoles completely cancel when all parameters are renormalized in the on-shell scheme.  However, QCD corrections to the $b$-quark mass and electric 
charge are sensitive to energy scales much smaller than the Higgs mass if the on-shell scheme is used,
so one would prefer to renormalize such parameters in the \msbar~scheme.  In that case tadpole 
cancellation can no longer occur, and tadpoles enter the finite parts of the renormalized
decay rate, carrying along with them corrections scaling as  $m_t^4/(\hat{v}_T^2 m_H^2)$,
which can lead to sizeable weak corrections. It is thus a non-trivial problem to find 
a renormalization scheme which is well suited for combining electroweak 
and QCD corrections in SMEFT.  We deal with this issue in the next section.

\section{Enhanced NLO corrections and decoupling relations}
\label{sec:DecouplingLargeCors}

The size of perturbative corrections to the decay rate depends on the renormalization scheme, 
and it is an important question whether it is possible to find a scheme which 
reduces the size of higher-order corrections. In section~\ref{sec:NLO_Structure} 
we identify sources of enhanced NLO corrections 
to the decay rate, and in section~\ref{sec:Decoupling} we emphasise the importance of 
decoupling particles with masses at the electroweak scale from the \msbar~definitions
of the $b$-quark mass and electric charge when combining QCD and electroweak corrections
in SMEFT.

\subsection{Structure of the NLO decay rate}
\label{sec:NLO_Structure}

The full NLO result for the decay rate, including mass dependence of third generation fermions, 
is quite lengthy.  However, it is possible to identify two sources of parametrically-enhanced corrections
and their dependence on the renormalization scheme.  
The first is logarithms of the small ratio $m_b/m_H$,  which appear in the QCD-QED type corrections
contained in the piece $\Gamma_{g,\gamma}$ defined  in  (\ref{eq:PertExpansion2}).  The 
result for these corrections in the $m_b\to 0$ limit is given in appendix~\ref{sec:Gqcd-qed}, using 
the notation in (\ref{eq:CXdef}) in order to keep the dependence on the renormalization
scheme for the $b$-quark mass explicit.  Setting $\mu=m_H$ and keeping only the logarithmic corrections in 
the result, one has
\begin{align}
\label{eq:LargeLogs}
\frac{\Gamma_{g,\gamma}^{(1)}}{\Gamma^{(4,0)}} \approx \, &   \ln^2\left(\frac{m_b^2}{m_H^2}\right) 
\frac{\hat{v}_T^2}{\pi}\left(C_F \alpha_s C_{HG}+ Q_b^2 \alpha c_{h\gamma\gamma}\right)
\nonumber \\ &
+\ c_{m_b}  \ln\left(\frac{m_b^2}{m_H^2}\right)  \frac{3}{2}\left(\frac{C_F \alpha_s+ Q_b^2 \alpha}{\pi}\right)
\bigg[1 + 2\hat{v}_T^2 \bigg(C_{H\Box}-\frac{C_{HD}}{4}\left(1- \frac{\hat{c}_w^2}{\hat{s_w^2}} \right)
\nonumber \\
&\hspace{6.58cm}+ \frac{\hat{c}_w}{\hat{s_w}} C_{HWB}  
-\frac{\hat{v}_T}{m_b} \frac{C_{bH}}{2\sqrt{2}}\bigg)\bigg] , 
\end{align}
where $c_{m_b} = 1$ ($c_{m_b} = 0$) yields the result in the on-shell scheme (\msbar~scheme)
for $m_b$.  It is simple to show that the decay rate in SMEFT depends only on the real parts of the Wilson 
coefficients, to the order which we are working.  We have therefore used the notation
that ${\rm Re}(C_i)\equiv C_i$ in writing (\ref{eq:LargeLogs}), and do this whenever we write an
expression for the decay rate in what follows.
Evaluating  (\ref{eq:LargeLogs}) numerically using the inputs in 
table~\ref{tab:Inputs} below yields 
\begin{align}
\label{eq:LargeLogsNum}
\frac{\Gamma_{g,\gamma}^{(1)}}{\Gamma^{(4,0)}} \approx \, &  
\hat{v}_T^2 \left(2.4 C_{HG}+  0.02  c_{h\gamma\gamma}\right) 
\nonumber \\* &
 -0.5c_{m_b}
\bigg[1 + 2\hat{v}_T^2 \bigg(C_{H\Box}-\frac{C_{HD}}{4}\left(1- \frac{\hat{c}_w^2}{\hat{s_w^2}} \right)
+ \frac{\hat{c}_w}{\hat{s_w}} C_{HWB}  
-\frac{\hat{v}_T}{m_b} \frac{C_{bH}}{2\sqrt{2}}\bigg)\bigg] \, .
\end{align}
We see that the QCD corrections are dominated by the double logarithmic term on the first line of (\ref{eq:LargeLogs}).  This term is of IR origin and  cannot be removed through a choice of renormalization scheme.\footnote{This contribution arises 
from the interference of the SM amplitude with dimension-6 amplitudes
involving $Hgg$ and $H\gamma\gamma$ vertices. These vertices do not contain a $b$-quark Yukawa coupling, so the fact that the  contribution to the decay rate scales as $m_b^2$ is  due to a chirality flip in the $b$-quark propagator,  which vanishes in the massless limit.  The appearance of this double logarithmic contribution 
is thus not in contradiction with the fact that the leading term in the limit $m_b\to 0$ should be  IR finite.}  It would need to be treated with QCD resummation techniques
which we do not explore here.  The single logarithmic term in the second 
and third line of (\ref{eq:LargeLogs}) arises from the finite part of the counterterm 
for $b$-quark mass renormalization in the on-shell scheme.
 Although not as large as the double logarithmic term,  it is still
a $-50\%$ correction to the LO result,  which can be removed from the explicit 
NLO correction and resummed by using the \msbar~scheme for the $b$-quark mass.
We conclude that the QCD-QED corrections to the decay rate
are best behaved in the \msbar~scheme for the $b$-quark mass, 
which is indeed standard in SM computations.  

The second source of potentially large corrections to the decay rate
are weak corrections enhanced by powers of $m_t^2/\hat{v}_T^2$, which
appear in the object $\Gamma_t$ defined in (\ref{eq:PertExpansion2}).
We give explicit results for the SM and dimension-6 corrections to
$\Gamma_t$ in appendix~\ref{sec:LMT}, as above using the notation in
(\ref{eq:CXdef}) in order to study the dependence on the
renormalization scheme.  The results show that in the \msbar~scheme
for the $b$-quark mass and electric charge the dominant contributions
are due to tadpoles and scale as $m_t^4/(\hat{v}_T^2m_H^2)$.  The
appearance of such corrections in NLO SMEFT calculations which make
use of the~\msbar~scheme has been emphasized in $h\to\gamma\gamma$
decay in~\cite{Hartmann:2015oia}, and in the partial NLO calculation
of $Z\to b\bar{b}$ in \cite{Hartmann:2016pil}.  In the on-shell scheme
tadpoles are absent and the leading corrections scale as $m_t^2/\hat{v}_T^2$.  We
translate this into numerical results using the SM as an example.
Keeping only the leading terms in the large-$m_t$ limit, one finds in
the \msbar~scheme
\begin{align}
\frac{\overline{\Gamma}_t^{(4,1)}}{\Gamma^{(4,0)}} &
\approx -\frac{N_c}{2\pi^2}\frac{m_t^4}{\hat{v}_T^2 m_H^2}\approx -15\% \, ,
\end{align}
 while in the on-shell scheme
 \begin{align}
\frac{\left[\Gamma_t\right]^{{\rm O.S.}(4,1)}}{\Gamma^{(4,0)}} & =
 \frac{m_t^2}{16 \pi^2 \hat{v}_T^2}\left(-6+ N_c \frac{7-10 \hat{c}_w^2}{3 \hat{s}_w^2}\right)  \approx -3\% \, ,
\end{align}
where we have set $\mu=m_t$ as appropriate in the large-$m_t$ limit 
and again used the inputs in table~\ref{tab:Inputs}.
The correction in the \msbar~scheme for the $b$-quark mass is
a $-15\%$ correction to the LO result and thus anomalously large for a 
weak correction, while that in the on-shell scheme takes on a much smaller
value, in line with naive expectations.  The numerical results for the dimension-6 contributions
differ from operator to operator, but it is still the case that the corrections tend to be 
larger in the \msbar~scheme than the on-shell one  due to tadpole corrections scaling as $m_t^4/(\hat{v}_T^2m_H^2)$.
 
The upshot of this discussion is that while the QED and QCD corrections are best 
behaved in the~\msbar~scheme for $m_b$, the electroweak
corrections are better behaved in the on-shell scheme for $m_b$ and $e$, where tadpole contributions from heavy particles such as the top quark cancel. 
At least in the SM, an apparent compromise would be to use the~\msbar~scheme for all parameters appearing in 
the tree-level result, be it quark masses, the electric charge, $M_W$ or $M_Z$.  This is however an imperfect solution, for although in that case no explicit tadpoles appear in the  NLO corrections, 
they reappear in the RG equations.  Moreover, in SMEFT it is not possible to remove all explicit tadpole contributions 
in this manner, since in contrast to the Standard Model they can also 
appear in the matrix elements for $h\rightarrow \bar{b}b$, through contributions such as that shown in Figure~\ref{fig:TadpoleAll}(d).

The resolution to this dilemma is to renormalize the $b$-quark mass and 
electric charge such that the QCD-QED corrections are treated in the \msbar~scheme, while weak corrections 
involving the top quark and heavy electroweak bosons are treated in the on-shell scheme. In that way contributions from
potentially large tadpole corrections cancel, but logarithms of $m_b/m_H$ can still be resummed in
the \msbar~scheme.   At the technical level, the simplest way to implement such a scheme is to make use of so-called "decoupling relations".

\subsection{Decoupling relations}
\label{sec:Decoupling}

Decoupling relations connect \msbar-renormalized parameters in SMEFT
with those defined in a low-energy theory where the top quark and
electroweak bosons are integrated out. A detailed discussion of this
in the SM for the $b$-quark mass defined in the \msbar~scheme can be
found in \cite{Bednyakov:2016onn}.  We shall consider only the
dimension-4 piece of this low-energy theory, which we refer to hereafter
simply as QED$\times$QCD.  This amounts to neglecting
terms which scale as e.g.  $m_b^2/M_W^2$, which are numerically
negligible compared to the dimension-4 terms.  We can then write the
decoupling relations as
\begin{align}
\label{eq:DecoupRelations}
\overline{m}_b(\mu)& = \zeta_b(\mu,m_t, m_H,M_W,M_Z) 
\overline{m}^{( \ell )}_b(\mu)  \, , \nonumber \\
\overline{e}(\mu) &= \zeta_e(\mu,m_t, m_H,M_W,M_Z) 
\overline{e}^{(\ell)}(\mu) \, ,
\end{align}
where the parameters on the left-hand side are defined in SMEFT, and those on 
right-hand side, with the superscript $\ell$, are defined in QED$\times$QCD.  
These parameters obey the RG equations
\begin{align}
\label{eq:5_flav_running}
\frac{d\overline{m}_b^{(\ell)}(\mu)}{d\ln\mu}&=\gamma_b(\mu) \,  
\overline{m}_b^{(\ell)}(\mu) \, ,
\nonumber \\
\frac{d\overline{e}^{(\ell)}(\mu)}{d\ln\mu}&=
\gamma_e(\mu)\,  \overline{e}^{(\ell)}(\mu) \,.
\end{align}
In what follows we will make use of the LO anomalous dimensions 
$\gamma_i$, which read
\begin{align}
\gamma_b(\mu) &=-\frac{3}{2\pi}\left[ \alpha_s(\mu) C_F + \overline{\alpha}^{(\ell)}(\mu) Q_b^2\right]  \, , 
\nonumber \\
\gamma_e(\mu) & = 
\frac{\overline{\alpha}^{(\ell)}(\mu)}{3\pi}\left[ N_g Q_\ell^2 +N_c \left((N_g-1)Q_u^2+N_g Q_b^2\right) \right]\,,
\label{eq:5_flav_anom}
\end{align}
where $N_g=3$ is the number of fermion generations, $Q_u=2/3$ for 
up-type quarks, and $\overline{\alpha}^{(\ell)}(\mu)\equiv[\overline{e}^{(\ell)}(\mu)]^2/(4\pi)$. The parameter $\overline{m}_b^{(\ell)}(\mu)$ is closely related to that used
in $B$ physics, where one typically includes only five-flavour QCD contributions to the running of  
$\overline{m}_b^{(\ell)}(\overline{m}_b^{(\ell)})\approx 4.2$~GeV.  On the 
other hand, the parameter $\overline{\alpha}^{(\ell)}(\mu)$ is related to the effective on-shell coupling 
$\alpha(M_Z)$ according to
\begin{align}
\frac{\overline{\alpha}^{(\ell)}(M_Z)}{\alpha(M_Z)}= 1+\frac{100\alpha}{27\pi} \, ,
\end{align}
where  $\alpha(M_Z)\approx 1/129$ compared the on-shell value $\alpha\approx 1/137$
(see e.g. \cite{Tanabashi:2018oca}).

The $\zeta_{i}$ in~eq.~(\ref{eq:DecoupRelations}) are decoupling constants.  They are determined by using the relation between
the \msbar~and on-shell parameters in the two theories.  These take the form
\begin{align}
m_b & = z_b^{-1}(\mu, m_b, m_t,m_H,M_W,M_Z)  \overline{m}_b(\mu) 
= \left[z^{(\ell)}_b(\mu, m_b)\right]^{-1}\overline{m}_b^{(\ell)}(\mu) \, ,  \nonumber \\
e &= z^{-1}_e(\mu, m_b, m_t,m_H,M_W,M_Z) \overline{e}(\mu) 
= \left[z_e^{(\ell)}(\mu, m_b)\right]^{-1}\overline{e}^{(\ell)}(\mu) \, , 
\end{align}
where we have used that the on-shell parameters $e$ and $m_b$ are 
defined through non-perturbative renormalization conditions and do not
depend on the Lagrangian. The $z_i$ factors are finite and determine the 
perturbative shifts between the on-shell and \msbar~parameters. 
They fix the decoupling constants through the relations
\begin{align}
\label{eq:DecDef}
\zeta_i(\mu,m_t, m_H,M_W,M_Z)  
= \frac{z_i(\mu, m_b, m_t,m_H,M_W,M_Z)}{z^{(\ell)}_i(\mu, m_b)} \bigg|_{m_b\to0}\,,
\end{align}
where $i=e,b$.   

We write the perturbative expansion of the decoupling constants in SMEFT as 
\begin{align}
\zeta_i= 1 +  \zeta_i^{(4,1)} + \zeta_i^{(6,1)} \, ,
\end{align}
where the superscripts $(4,1)$ and $(6,1)$ follow the notation of
(\ref{eq:PertExpansionSMEFT}).  At NLO the decoupling constants
are proportional to the finite parts of heavy-particle contributions
to the NLO renormalization constants.  The expression for $\zeta_e$ is
compact.  The SM expression is
\begin{align}
\zeta_e^{(4,1)}= \frac{\alpha}{ \pi} \left[-\frac{1}{12} -\frac{7}{8} \ln \left( \frac{\mu^2}{M_W^2} \right)
+\frac{N_c}{6}Q_t^2  \ln \left( \frac{\mu^2}{m_t^2}\right)\right] \, , 
\end{align}
and the SMEFT result reads  
\begin{align}
\label{eq:ze6}
\zeta_e^{(6,1)} =&
\frac{\alpha}{\pi}\left[\sqrt{2}\hat{v}_T m_t N_c Q_t
 \left(\hat{c}_w \frac{{\rm Re}(C_{tB})}{e}+\hat{s}_w \frac{{\rm Re}(C_{tW})}{e}\right)   \ln \left( \frac{\mu^2}{m_t^2} \right)
+ 9 \frac{C_W}{e} \hat{s}_w  M_W^2  \ln\left( \frac{\mu^2}{M_W^2}\right)
\right]
	\nonumber \\ 
	&+ \frac{\delta e^{{\rm cl.4}(6)}}{e} \bigg|_{{\rm fin.,} \, m_b \to 0}  \, ,
\end{align}
where $Q_t=2/3$ is the charge of the top quark and the term on the second line of (\ref{eq:ze6}) 
is the UV-finite  part of the class-4 electric charge counterterm (\ref{eq:deltaEC4Feyn}) with $m_b\to 0$.  

The full results for $\zeta_b$ are somewhat lengthy, and are given in computer files in the arXiv version of this paper.
They simplify considerably in the large-$m_t$ limit, where they read 
\begin{align}
\zeta_{b,t}^{(4,1)} =  \delta b_t^{(4)}, \qquad \zeta_{b,t}^{(6,1)} =  \delta b_t^{(6)} \, , 
\end{align}
where $\delta b_t$ is the UV-finite part of  $\delta m_b$ in this limit and is given in (\ref{eq:DBT}).

The $h\to b\bar{b}$ decay rate written in terms of the QCD$\times$QED parameters 
$\overline{m}_b^{(\ell)}$ and  $\overline{e}^{(\ell)}$, which we denote by $\overline{\Gamma}_\ell$, 
is simple to obtain from the decay  rate in terms of the parameters $\overline{m}_b$ and  $\overline{e}$ in the 
full SMEFT, which we denote by  $\overline{\Gamma}$. 
The LO results are the same up to a renaming of the parameters, and the NLO results are given by
\begin{align}
\label{eq:GammaDec}
\overline{\Gamma}_\ell^{(4,1)} &=\overline{\Gamma}^{(4,1)} +
2 \overline{\Gamma}^{(4,0)}\left(\zeta_b^{(4,1)}+\zeta_e^{(4,1)}\right)
 \, , \nonumber \\
 \overline{\Gamma}_\ell^{(6,1)}  &=  \overline{\Gamma}^{(6,1)} +
 2\overline{\Gamma}^{(4,0)}\left(\zeta_b^{(6,1)}+\zeta_e^{(6,1)}\right)
 +2\overline{\Gamma}^{(6,0)} \zeta_b^{(4,1)} 
 \nonumber \\*
 &+ \sqrt{2} C_{bH}\frac{(\overline{v}^{(\ell)})^3}{\overline{m}^{(\ell)}_b} \overline{\Gamma}^{(4,0)}  \left( \zeta_b^{(4,1)} +\zeta_e^{(4,1)} \right) \,,
\end{align}
where  we have suppressed  dependence on the \msbar~renormalization scale $\mu$ and 
introduced   
\begin{align}
\overline{v}^{(\ell)}(\mu)\equiv \frac{2M_W \hat{s}_w}{\overline{e}^{(\ell)}(\mu)}\,.
\end{align}
Eq.~(\ref{eq:GammaDec}) is obtained by inserting (\ref{eq:DecoupRelations}) into $\overline{\Gamma}^{(0)}$ 
and expanding to NLO.  The same result can be obtained by replacing 
$\delta m_b/m_b\to \delta m_b/m_b+\zeta_b$ and similarly for $\delta e/e$ in the NLO 
counterterms (\ref{eq:dim4CT}) and (\ref{eq:dim6CT}), and for this reason evaluating the
decay rate using (\ref{eq:GammaDec}) is equivalent to using a  new renormalization scheme.  
After splitting up the decay rate in this scheme as
\begin{align}
 \overline\Gamma_\ell^{(1)} = 
 \overline{\Gamma}^{(1)}_{\ell, g,\gamma}+\overline{\Gamma}^{(1)}_{\ell,t} + 
 \overline{\Gamma}^{(1)}_{\ell,\rm rem} \,,
\end{align}
it is possible to list a simple and illustrative result for the QCD$\times$QED and large-$m_t$ limit of the 
weak corrections.  In terms of the quantities defined in appendix~\ref{sec:analytic}, we have
\begin{align}
\label{eq:Gam5}
\overline{\Gamma}_{\ell, g,\gamma} = \overline{\Gamma}_{g,\gamma} \, ,
\qquad \overline{\Gamma}_{\ell,t}= \left[\Gamma_t\right]^{\rm O.S.} \, .
\end{align}
The interpretation is that the QCD$\times$QED corrections are calculated in 
the \msbar~scheme, while contributions from top-quark loops are 
calculated in the on-shell scheme, where tadpoles cancel.  
This pattern holds for heavy gauge-boson contributions to the decay
rate.    In fact,  after decoupling, heavy-particle contributions are effectively
calculated in the on-shell scheme, so that the only non-vanishing tadpole contributions 
are suppressed by powers of light fermion masses and are negligible numerically.

\section{Numerical results}
\label{sec:Results}
In this section we present results for the $h\to b\bar{b}$ decay rate
at NLO in SMEFT.  We first give numerical results with the default
choice $\mu = m_H$ in section~\ref{sec:DefaultGamma}, and then perform
a study of perturbative uncertainties due to scale variations in
section~\ref{sec:Uncertainties}.  Throughout the analysis we use the
renormalization scheme defined in (\ref{eq:GammaDec}).  Since the
decoupling relations used in that scheme are
valid in the limit where all fermion masses except the top-quark mass
$m_t$ vanish, we shall use this approximation in presenting the
numerical results. The dominant corrections to this limit scale as 
$m_b^2/M_W^2$ and typically change the NLO {\it corrections} at the
1\% level and are thus irrelevant for our discussion.  The input
parameters needed in the analysis are listed in
table~\ref{tab:Inputs}.

\subsection{Results at $\mu=m_H$}
\label{sec:DefaultGamma}

To quote results for the dimension-6 contributions, we make the dependence on $\Lambda_{\rm NP}$ 
explicit by defining dimensionless Wilson coefficients according to 
\begin{align}
  \tilde{C}_i(\mu)  \equiv  \Lambda_{\rm NP}^2 C_i(\mu)  \,.
\end{align} 
Contributions to the decay rate from dimension-6 operators are then 
suppressed by an explicit power of  $\bar{v}^{(\ell)}(\mu)^2/\Lambda_{\rm NP}^2$, which for the input 
parameters in table~\ref{tab:Inputs}  leads to a roughly 5\% suppression factor for
$\Lambda_{\rm NP}=1$~TeV and $\tilde{C}_i\sim 1$.   
\begin{table}[t]
	\begin{center}
		\def\arraystretch{1.3}
		\begin{tabular}{|c|c||c|c|}
			\hline  $m_H$ &  $125$~GeV   & $\overline{m}_b^{(\ell)}(m_H)$ & $3.0$~GeV \\ 
			\hline $m_t$ & $173$~GeV& $\overline{e}^{(\ell)}(m_H)$ & $\sqrt{4 \pi/ 128}$ \\ 
			\hline $M_W$ & $80.4$~GeV &  $\overline{v}^{(\ell)}(m_H)$ & 240 GeV \\ 
			\hline $M_Z$ & $91.2$~GeV &$\alpha_s \left(m_H\right)$ & 0.1 \\
			\hline 
		\end{tabular} 
		\caption{\label{tab:Inputs}Input parameters employed throughout the calculation, where 
		we have also listed the derived quantity $\overline{v}^{(\ell)}(m_H) \equiv 2M_W \hat{s}_w/\overline{e}^{(\ell)}(m_H)$ 
		for convenience.}
	\end{center}
\end{table}

We shall present numerical results normalized to the LO SM decay rate.  We thus define
\begin{align}
\Delta^{\rm LO}(\mu) &\equiv 
 \frac{\overline{\Gamma}_\ell^{(4,0)}(\mu)+\overline{\Gamma}_\ell^{(6,0)}(\mu)}{\overline{\Gamma}_\ell^{(4,0)}(m_H) } \, , 
 \nonumber \\
 \Delta^{\rm NLO}(\mu)& \equiv 
 \Delta^{\rm LO}(\mu)+\frac{\overline{\Gamma}_\ell^{(4,1)}(\mu)+\overline{\Gamma}_\ell^{(6,1)}(\mu)}{\overline{\Gamma}_\ell^{(4,0)}(m_H)}   \,.
\end{align}
Using $\mu=m_H$ and supressing the arguments on $\overline{v}^{(\ell)}(m_H)$ and 
$\tilde{C}_i(m_H)$,  we find 
\begin{align} 
\Delta^{\rm LO}(m_H) =
 1+\frac{(\overline{v}^{(\ell)})^2}{\Lambda_{\rm NP}^2}
 \left[
 3.74  \Ct_{HWB} + 2.00 \Ct_{H\Box}-1.41 \frac{\bar{v}^{(\ell)}}{\overline{m}^{(\ell)}_b} \Ct_{bH}+1.24 \Ct_{HD}
 \right] \, .
\end{align}
In quoting this result, we have kept a factor of $\bar{v}/m_b\sim 80$ multiplying the $\tilde{C}_{bH}$ contribution
symbolic.  We do this to highlight the fact that the $\tilde{C}_{bH}$ contribution to the decay rate 
scales as $m_b$ rather than $m_b^2$ as in the SM, which can be seen explicitly in (\ref{eq:LOgam}).  
The same is true of six additional coefficients which enter the decay rate at NLO:  
$\tilde{C}_{bG}$, $\tilde{C}_{bW}$, $\tilde{C}_{bB}$, $\tilde{C}_{Htb}$, 
$\tilde{C}^{(1)}_{qtqb}$ and $\tilde{C}^{(8)}_{qtqb}$.  It is worth mentioning that if
MFV is imposed then all of these coefficients scale as $y_b\sim m_b/\bar{v}$, so that their contributions to the
decay rate scale as $m_b^2$. However, our results are not limited to MFV, 
so keeping factors of $\bar{v}/m_b$ symbolic when multiplying the coefficients mentioned above is simply a 
matter of convenience.  For the same
reason, when quoting results from operators such as $Q_{bB}$ or $Q_{bG}$ 
where gauge bosons couple through field strengths rather than covariant derivatives,  
we keep enhancement factors of $1/\overline{e}$ or $1/g_s$ compared to the SM contributions symbolic.  With these
conventions, the NLO result can be written as 
\begin{align} 
\Delta^{\rm NLO}&(m_H)  = 1.13 + 
 \frac{(\bar{v}^{(\ell)})^2}{\Lambda_{\rm NP}^2}
 \bigg\{
 4.16  \Ct_{HWB} +
2.40 \Ct_{H\Box} - 1.73 \frac{\bar{v}^{(\ell)}}{\overline{m}_b^{(\ell)}} \Ct_{bH}+1.33 \Ct_{HD}
 \nonumber \\ & 
 + 2.75 \Ct_{HG}- 0.12 \Ct_{Hq}^{(3)}+
\bigg(-7.9 \Ct_{Ht} + 5.8 \Ct_{Hq}^{(1)} + 3.1 \frac{\bar{v}^{(\ell)}}{\overline{m}_b^{(\ell)}} \Ct_{qtqb}^{(1)}
-3.1 \Ct_{tH}+2.7 \Ct_{HW}
 \nonumber \\ &
 +2.4 \Ct_{H}-1.9\frac{\bar{v}^{(\ell)}}{\overline{e}^{(\ell)}\,\overline{m}_b^{(\ell)}}\Ct_{bW}-1.3 \Ct_{qb}^{(8)}
-1.3 \frac{\Ct_{tW}}{\overline{e}^{(\ell)}} - 1.0 \Ct_{qb}^{(1)}
\bigg)\times10^{-2}  \nonumber \\
&+ \bigg(-9\left[\frac{\Ct_{tB}}{\overline{e}^{(\ell)}}+\left(\Ct_{Hq}^{(3)}\right)_{22}
+ \left(\Ct_{Hq}^{(3)}\right)_{11}- \Ct_{HB} + \Ct_{Hu}+\Ct_{Hc}\right]
- 8 \frac{\bar{v}^{(\ell)}}{g_s \overline{m}_b^{(\ell)}} \Ct_{bG} 
- 7 \Ct_{W}  \nonumber \\
&+ 6 \frac{\bar{v}^{(\ell)}}{\overline{m}_b^{(\ell)}} \Ct_{qtqb}^{(8)}
+4\bigg[ \Ct_{Hl}^{(1)}+
\left(\Ct_{Hl}^{(1)}-\Ct_{Hq}^{(1)}\right)_{22}
+\left(\Ct_{Hl}^{(1)}-\Ct_{Hq}^{(1)}\right)_{11}
+\Ct_{H\tau}+\Ct_{H\mu}+\Ct_{He}\nonumber \\
& +\Ct_{Hs}+\Ct_{Hd}-\frac{\bar{v}^{(\ell)}}{\overline{m}_b^{(\ell)}} \Ct_{Htb}
 \bigg] - 3\bigg[  \Ct_{Hl}^{(3)}+
 \left(\Ct_{Hl}^{(3)}\right)_{22}+
 \left(\Ct_{Hl}^{(3)}\right)_{11}
\bigg]+2 \Ct_{Hb} 
   \bigg)\times 10^{-3}  \nonumber \\
   &- 4\times 10^{-5} \frac{\bar{v}^{(\ell)}}{\overline{e}^{(\ell)} \, \overline{m}_b^{(\ell)}}\Ct_{bB} 
 \bigg\} \, .
\end{align}
By far the largest NLO correction is from $\tilde{C}_{HG}$, which is a QCD
effect enhanced by a double logarithm in $m_b/m_H$ as described in
section~\ref{sec:NLO_Structure}.  Order 10\% corrections (in units of
$\bar{v}^2/\Lambda_{\rm NP}^2$) arise from $\Ct_{Hq}^{(1)}$,
$\Ct_{Hq}^{(3)}$ and $\Ct_{Ht}$.  In total there are 16 operators
which contribute at greater than a percent level to the decay rate, 12
of which first appear at NLO.

\begin{table}[t]
\begin{center}
\begin{tabular}{l|rrrrr}
 & SM  & $\Ct_{HWB}$
& $\Ct_{H\Box}$  & $\Ct_{bH}$  & $\Ct_{HD}$ \\
\hline 
NLO QCD-QED  & 18.2\% & 17.9\% & 18.2\%  &18.2\% &  18.2\%   \\
NLO large-$m_t$ & -3.1\% & -4.6\% &  3.2\%  & 3.5\% &  -9.0\%   \\
NLO remainder & -2.2\% & -1.9\% &  -1.2 \%  & 0.6\% &  -2.0\%   \\
\hline 
NLO correction & 12.9\% & 11.3\% &  20.2\%  &22.3\% &  7.1\%   \\ 
\end{tabular}
\caption{\label{tab:CorSplit} 
Size of NLO corrections to different terms in LO decay rate, split into
QCD-QED, large $m_t$, and remaining components.  See text for further explanation.  }
\end{center}
\end{table}

Generally speaking, an operator gives a significant contribution only
if it involves QCD or large-$m_t$ corrections.  To illustrate the
relative importance of these two effects, we show in
table~\ref{tab:CorSplit} the division of the NLO corrections to
operators appearing at tree level into QCD-QED corrections,
large-$m_t$ corrections, and remaining corrections (denoted by
$\overline{\Gamma}_{\ell, g,\gamma}$, $\overline{\Gamma}_{\ell,t}$, and
$\overline{\Gamma}_{\ell,\rm rem}$).  For the dimension-6 operators, the
numbers are defined as the contribution of the Wilson coefficient
$\tilde{C}_i$ to $\overline{\Gamma}_\ell^{(1)}$ divided by its contribution to
$\overline{\Gamma}_\ell^{(0)}$.  The results show that while the QCD
corrections are dominant, the electroweak corrections are
non-negligible and depend strongly on the Wilson coefficient.  For
instance, the electroweak corrections from $\tilde{C}_{HD}$ are $-11$\%, while
those from $\tilde{C}_{bH}$ are $+3\%$. Therefore, approximating the NLO
corrections in SMEFT by multiplying the tree level result with a
universal $K$-factor derived from the SM QCD corrections would be a
poor estimate to the full calculation performed here. We also note
that the large-$m_t$ corrections indeed make up the bulk of the electroweak
corrections, although deviations from that approximation are between
$10-40\%$. We have observed that this pattern holds for the other
coefficients appearing in the NLO result.

\subsection{Scale uncertainties}
\label{sec:Uncertainties}
So far we have given results only at $\mu= m_H$.  In this 
section we address two obvious questions concerning 
scale uncertainties: first, can the size of NLO corrections
be reliably estimated through scale variations of the LO result, and second, 
what is the residual uncertainty beyond NLO?  

We shall study these questions as typical in a perturbative analysis, namely by varying unphysical 
renormalization scales up and down by factors of two and taking the change in the decay rate as 
a measure of the uncertainty due to uncalculated, higher-order corrections.
A difference in SMEFT compared to the  SM is that while all parameters in the 
SM Lagrangian have been determined to good accuracy numerically, the exact values of the 
Wilson coefficients in SMEFT are largely unknown. Therefore, when performing scale variations, 
we give results symbolically in terms of the Wilson coefficients at a fixed reference scale. In our case, the natural choice of this reference scale is $\mu=m_H$, therefore our task is 
to express the Wilson coefficients $C_i(\mu)$ in terms of the $C_i(m_H)$.  This is achieved by 
solving the RG equations for the Wilson coefficients. 

For variations of $\mu$ by factors of two, $\mu\sim m_H$
parametrically, so we can use the fixed-order expansion of the RG
equations rather than the exact, exponentiated solution.  In fact, the
same holds for the SM masses and couplings renormalized in the
\msbar~scheme.  Given that the anomalous dimensions of the Wilson
coefficients are known only to one-loop, we use this same level of
accuracy for the SM parameters throughout this section.  The solutions
of the RG equations to NLO in fixed order read
\begin{align}
\label{eq:Cevolve1}
C_i(\mu_C) & = C_i(m_H) +\ln\left(\frac{\mu_C}{m_H}\right)
 \dot{C}_i(m_H) \,  ,
\nonumber \\
\overline{m}_b^{(\ell)}(\mu_R)&=\overline{m}_b^{(\ell)}(m_H)
\left[1+ \gamma_b(m_H) \ln\left(\frac{\mu_R}{m_H}\right)\right] ,
\nonumber \\
\overline{\alpha}^{(\ell)}(\mu_R) & = \overline{\alpha}^{(\ell)}(m_H) \left[ 1+2 \gamma_e(m_H) \ln\left(\frac{\mu_R}{m_H}\right) \right] \, , \nonumber \\
\alpha_s(\mu_R) & = \alpha_s(m_H)\left[ 1-2 \gamma_g(m_H) \ln\left(\frac{\mu_R}{m_H}\right) \right] \, ,
\end{align}
where 
\begin{align}
\gamma_g(\mu_R)= \frac{\alpha_s(\mu_R)}{4\pi} \left( \frac{11}{3} C_A-\frac{2}{3}n_l\right)   \, .
\end{align}
The number of light quarks is $n_l=5$ and $C_A=3$. Results for $\gamma_e$ and $\gamma_b$ were 
given in (\ref{eq:5_flav_anom}), and $\dot{C}_i$ was defined in (\ref{eq:delCdef}).

We have written (\ref{eq:Cevolve1}) in a fashion which emphasizes that it is possible to use different renormalization scales $\mu_C$ and $\mu_R$ for the Wilson 
coefficients  and the SM parameters, respectively.  Until this point we have
set $\mu_C=\mu_R=\mu$, but in our scale uncertainty analysis it will be useful to
consider independent variations of these scales.  
These scales appear not only implicitly in the Wilson coefficients, $b$-quark mass, 
and the strong and electromagnetic coupling constants,  but also in explicit logarithms in the NLO decay rate. The explicit logarithmic dependence on the two  scales in the NLO  dimension-6 results can
be  reconstructed from the result at $\mu_R=\mu_C = \mu$ by using the RG equations  
along with the requirement that the decay rate is independent of the renormalization 
scales up to terms of order NNLO and higher.  The results can be written 
as 
\begin{align}
\label{eq:LogConvert}
\overline{\Gamma}_\ell^{(6,0)}(\mu_R, \mu_C)&= \overline{\Gamma}_\ell^{(6,0)}(\mu_C) \bigg
|_{\overline{p}(\mu_C)\to  \overline{p}(\mu_R)} \, , \nonumber \\
\overline{\Gamma}_\ell^{(6,1)}(\mu_R, \mu_C)& =\bigg\{ \overline{\Gamma}_\ell^{(6,1)}(\mu_C)  
+  2\left[ \ln \left( \frac{\mu_C}{m_H} \right)- \ln \left( \frac{\mu_R}{m_H} \right) \right] \bigg(\gamma_b(\mu_C) \overline{\Gamma}_\ell^{(6,0)}(\mu_C)   \notag \\ 
&+   \frac{C_{bH}(\mu_C)}{\sqrt{2}}  \frac{(\overline{v}^{(\ell)})^3(\mu_C)}{\overline{m}^{(\ell)}_b(\mu_C)} \overline{\Gamma}_\ell^{(4,0)}(\mu_C) \big[\gamma_b(\mu_C)+\gamma_e(\mu_C)\big] \bigg)\bigg\}\bigg 
|_{\overline{p}(\mu_C)\to  \overline{p}(\mu_R)} \, ,
\end{align}
where $\overline{p}(\mu)\in  \{\overline{\alpha}^{(\ell)}(\mu), \, \overline{m}_b^{(\ell)}(\mu), \,   \alpha_s(\mu)\}$ 
are the \msbar-renormalized parameters appearing in the calculation. By definition 
$\overline{\Gamma}_\ell^{(6,i)}(\mu,  \mu)=\overline{\Gamma}_\ell^{(6,i)}(\mu)$.

With these pieces at hand, we obtain scale uncertainties using the following procedure.
For the SM results, we vary the scale $\mu_R$ up and down around its default value $m_H$.
For  the dimension-6 results,  we can vary both $\mu_R$ and $\mu_C$ using (\ref{eq:LogConvert}). 
The default setting is $\mu_R=\mu_C=m_H$.   We then assign an uncertainty to each scale
individually by varying it up and down by a factor of two while leaving the other scale fixed, and add 
the resulting uncertainties from the independent $\mu_R$ and $\mu_C$ variations in 
quadrature to obtain a total uncertainty.  The numerical values of the 
scale-dependent parameters at the different scales are determined in terms of their values at $m_H$ using (\ref{eq:Cevolve1}).
This results in
\begin{align} 
\label{eq:LO_Quad}
\Delta^{\rm LO}&(m_H,m_H)  = (1\pm 0.08) 
+ \frac{(\bar{v}^{(\ell)})^2}{\Lambda_{\rm NP}^2}
 \bigg\{
 \nonumber \\ 
 &
  (3.74 \pm 0.36)  \Ct_{HWB} + (2.00 \pm 0.21) \Ct_{H\Box}-(1.41\pm 0.07) \frac{\bar{v}^{(\ell)}}{\overline{m}^{(\ell)}_b} \Ct_{bH} 
   + (1.24 \pm 0.14) \Ct_{HD} \nonumber \\ & 
\pm 0.35 \Ct_{HG} \pm 0.19 \Ct_{Hq}^{(1)}  \pm 0.18 \Ct_{Ht}
 \pm 0.11 \Ct_{Hq}^{(3)} \nonumber \\
 &
\pm 0.08 \frac{\bar{v}^{(\ell)}}{\overline{m}^{(\ell)}_b} \Ct_{qtqb}^{(1)} \pm  0.03 \frac{\Ct_{tW}}{\overline{e}^{(\ell)}}
\pm 0.03 (\Ct_{HW}+\Ct_{tH}) + \dots \bigg\} \, , 
\end{align}
where the ellipses indicate dimension-6 terms which contribute less than 3\% 
in units of $\bar{v}^2/\Lambda_{\rm NP}^2$. At NLO, we find 
\begin{align} 
\label{eq:NLO_Quad}
\Delta^{\rm NLO}&(m_H,m_H)  = 1.13^{+0.01}_{-0.04} + 
 \frac{(\bar{v}^{(\ell)})^2}{\Lambda_{\rm NP}^2}
 \bigg\{
 \left(4.16^{+0.05}_{-0.14} \right) \,  \Ct_{HWB} +
 \left(2.40^{+0.04}_{-0.09}\right)\, \Ct_{H\Box} 
\nonumber \\* &
+\left(-1.73^{+0.04}_{-0.03}\right)\, \frac{\bar{v}^{(\ell)}}{\overline{m}^{(\ell)}_b} \Ct_{bH}
+\left(1.33^{+0.01}_{-0.04} \right)\Ct_{HD}
 +\left(2.75^{+0.49}_{-0.48} \right) \Ct_{HG}
  \nonumber \\* & 
+ \left( -0.12^{+0.04}_{-0.01} \right) \, \Ct_{Hq}^{(3)}
+ \left(-0.08^{+0.05}_{-0.01} \right) \,  \Ct_{Ht}
+\left(0.06^{+0.00}_{-0.05} \right)\,\Ct_{Hq}^{(1)} 
\nonumber \\* &
+\left(0.03^{+0.02}_{-0.01} \right)\, \frac{\bar{v}^{(\ell)}}{\overline{m}^{(\ell)}_b} \Ct_{qtqb}^{(1)}
+\left(0.00^{+0.07}_{-0.04}\right) \, \frac{\Ct_{tG}}{g_s} + \left(-0.03^{+0.01}_{-0.01}\right)\Ct_{tH} \nonumber \\ &
+\left(0.03^{+0.01}_{-0.01}\right)\Ct_{HW}+\left(-0.01^{+0.01}_{-0.00}\right)\Ct_{tW} + \ldots
 \bigg\} \, .
\end{align}
where now the ellipses indicate terms with  {\it uncertainties} smaller than 3\%, other than those which appear already in eq.~(\ref{eq:LO_Quad}).

We see that the NLO calculation generally leads to a 
considerable reduction in the scale uncertainties compared to LO. 
For the operators already appearing at tree level, the NLO corrections are on the
upper limits of what one would estimate through scale variations of the LO 
result.  For operators which first appear at NLO, varying the scale in the
LO results generally estimates the size of the NLO contribution quite well.
A major exception is the $\tilde{C}_{HG}$ coefficient.  In that case the size
of the NLO correction is dramatically underestimated by scale variations in 
the LO result, and in fact the NLO result has a larger perturbative 
uncertainty associated with it than the leading one.  This is not surprising, 
given that the large correction from $\tilde{C}_{HG}$ is completely unrelated to RG 
running, as explained in section~\ref{sec:NLO_Structure}.  A consequence of this is that a new coefficient
$\tilde{C}_{tG}$, which arises predominantly
through the running of $\tilde{C}_{HG}$, is a significant source of uncertainty in the NLO calculation.

Needless to say, the uncertainties assigned to the decay rate through the above
procedure are just estimates, and other methods for varying the scales
are possible.  The simplest one is to set $\mu_R=\mu_C=\mu$ and obtain 
uncertainties by varying the single 
scale $\mu$ up and down by a factor of two. Analytic results for the
uncertainties in the LO result, which we denote by  $\delta \overline{\Gamma}_\ell^{(i,0)}$,  
obtained in this way are quite simple:  dropping terms of order NNLO and higher, one has
\begin{align}
\label{eq:LO_Lin_Analytic}
\delta \overline{\Gamma}_\ell^{(4,0)} &= \pm 2\ln(2)\overline{\Gamma}_\ell^{(4,0)}\left(\gamma_b+\gamma_e\right)
 \, , \nonumber \\
\delta \overline{\Gamma}_\ell^{(6,0)} &= \pm 2\ln(2)\left[ \gamma_b \overline{\Gamma}_\ell^{(6,0)}+\frac{C_{bH}(\bar{v}^{(\ell)})^2}{\sqrt{2}}\frac{\bar{v}^{(\ell)}}{\overline{m}^{(\ell)}_b}  \overline{\Gamma}_\ell^{(4,0)} 
\left( \gamma_b+\gamma_e \right)+\frac{1}{2}\overline{\Gamma}_\ell^{(6,0)}\bigg|_{C_i\to \dot{C}_i}  \, \right] \, ,
\end{align}
where all scale-dependent quantities are to be evaluated at $\mu=m_H$. Compared to the results 
(\ref{eq:LO_Quad}) using the quadrature method, only contributions from the dimension-6 coefficients  
appearing in the LO matrix elements are changed.  Numerically evaluating (\ref{eq:LO_Lin_Analytic})
leads to the following result for those coefficients in units of $\bar{v}^2/\Lambda_{\rm NP}^2$:
\begin{align} 
 (3.74 \pm 0.20)  \Ct_{HWB} + (2.00 \pm 0.06) \Ct_{H\Box}-(1.41\pm 0.08) \frac{\bar{v}^{(\ell)}}{\overline{m}^{(\ell)}_b} \Ct_{bH} 
 + (1.24 \pm 0.02) \Ct_{HD}  \,.
\end{align}
The result for  $\Ct_{bH}$  is almost identical to that obtained
with the quadrature method, but the uncertainties assigned to the other coefficients are significantly smaller.  Especially those for $\Ct_{H\Box}$ (3\%) and $\Ct_{HD}$ (2\%) are artificially small uncertainties
to assign to an LO calculation, and for this reason we have chosen the quadrature method by default.

Even more conservative methods could be used, for instance a scan over
$\mu_C$ and $\mu_R$ which takes into account simultaneous but
uncorrelated variations to include choices such as $\mu_R=m_H/2,
\mu_C=2m_H$ where neither scale is at its default value, but we do not
explore such options here.  Our main message is that it is important
to assign uncertainties to the LO result, and these uncertainties are
significantly reduced through the NLO calculation.

\section{Conclusions}
\label{sec:conclusions}

We have calculated the full set of NLO corrections to $h\to b\bar{b}$ decay
in SMEFT, obtaining  contributions from the 45 dimension-6 Wilson
coefficients which enter the decay rate at this order.  These results form the 
basis for any future precision analysis of this decay in effective field theory.  
\nolinebreak
While the renormalization of the electroweak sector of
SMEFT is conceptually similar to the SM, in
section~\ref{sec:Renormalization} we highlighted some 
technical differences regarding charge renormalization and also Higgs mixing with the $Z$ and
neutral Goldstone bosons. Moreover, the structure of tadpole
cancellation in the $h\to b\bar{b}$ decay amplitude in the on-shell
renormalization scheme is rather intricate in SMEFT, since contrary to
the SM, tadpole contributions to the matrix elements, $b$-quark
wavefunction renormalization and electric charge renormalization must
be taken into account. 

Our calculation includes both electroweak and QCD corrections, which 
has led us to explore hybrid renormalization schemes
where heavy particle masses are renormalized on-shell while the 
$b$-quark mass and electric charge are renormalized in the
\msbar~scheme. In such schemes tadpoles do not cancel from the decay
amplitude, need to be included in order to obtain gauge invariant
decay results, and can lead to enhanced electroweak corrections.  In
section~\ref{sec:DecouplingLargeCors} we showed how these enhanced
electroweak corrections can be removed from the decay rate by
decoupling contributions from electroweak-scale masses from the
running of \msbar~renormalized parameters, which are then defined in a
low-energy version of QED$\times$QCD.  We obtained the
decoupling constants for the electric charge and $b$-quark mass 
to NLO in SMEFT, and used them to calculate the decay 
rates in a hybrid renormalization scheme which simultaneously avoids
enhanced tadpoles corrections from the electroweak sector and resums UV
logarithms in $m_b/m_H$ in the QCD one. 

In section~\ref{sec:Results} we gave numerical results in the aforementioned 
renormalization scheme with the scale choice $\mu=m_H$ for all  \msbar-renormalized
parameters, namely the Wilson coefficients as well as the $b$-quark mass and electric charge.
We also studied the perturbative uncertainties in the LO and NLO results as estimated through 
scale variations. We found that while in general the NLO corrections stabilize the scale dependence of 
the decay rate, genuine NLO effects inaccessible to an RG analysis based on scale variations can be significant. 
That said, we advocated introducing two renormalization scales, one for the 
Wilson coefficients and one for the \msbar~renormalized
$b$-quark mass and electric charge, and varying them independently in order to generate more reliable uncertainty
estimates than those obtained from varying a common scale $\mu$ .

The analytic results for the NLO decay rate in SMEFT are rather lengthy and included in computer files 
with the arXiv submission of this article, both with the full $m_b$ dependence, which will be useful for 
future validations of our results, and in the $m_b\to 0$ limit, which is sufficient for phenomenology.  We believe
that the renormalization procedure and uncertainty analysis performed here can serve as a template for 
future NLO SMEFT calculations which aim to include  electroweak and QCD corrections in a single
framework.

\section*{Acknowledgements}
The research of J.M.C.~is supported by an STFC Postgraduate Studentship. D.J.S.~is supported under the ERC grant ERC-STG2015-677323. The authors are grateful to Rhorry Gauld for collaboration on early stages of this work.

\appendix

\section{SMEFT in the mass basis}
\label{sec:MassBasis}

In the following sections we give some details on writing the SMEFT Lagrangian in the mass basis after
EWSB.  The discussion closely follows that in  \cite{Alonso:2013hga}, and our main goal is to keep track
of dimension-6 effects related to expressing the Lagrangian in terms of the physical observables in
(\ref{eq:InputPar}).  

\subsection{The Higgs doublet, vacuum expectation value and mass}
\label{sec:MassBasisHiggs}

The class-2 operator $C_H$ alters the SM expression for the vacuum expectation value of the Higgs field.
Defining the Higgs potential in the SM as
\begin{align}
V^{\rm SM}(H)=\lambda(H^\dagger H- v^2/2)^2 \,,
\end{align}
one finds that the vacuum expectation value is shifted  by dimension-6 corrections
from the SM value $v$ according to
\begin{align}
\langle H^\dagger H \rangle \equiv \frac{1}{2}v_T^2 = 
\frac{v^2}{2} \left(1+ \frac{ 3 C_H \hat{v}_T^2}{4\lambda }\right) \, .
\end{align}

Class 3 introduces operators that contribute to the kinetic terms of fields found in the Higgs doublet, these being the Higgs field and the neutral and charged Goldstone bosons. Appropriate field redefinitions must be made to restore the canonical normalization of the kinetic terms. As a result the Higgs doublet is written in Feynman gauge as
\begin{equation}
\label{eq:HiggsDoublet}
H(x)= \frac{1}{\sqrt{2}} 
\left( {\begin{array}{cc}
	- \sqrt{2} i \phi^+ (x) &\\
	\left[ 1+ C_{H,\text{kin.}} \right] h(x) + i\left[1-\frac{\hat{v}_T^2}{4}C_{HD}  \right]\phi^0(x) +v_T
	\end{array}}\right) \, ,
\end{equation}
where we have defined
\begin{equation}
\label{eq:CHkin}
C_{H,\rm kin}\equiv \left( C_{H \Box}-\frac{1}{4} C_{HD}\right)\hat{v}_T^2 \, .
\end{equation}
Notice that in the equations above we have replaced $v_T$ with $\hat{v}_T$ defined in (\ref{eq:SMVar}) 
when it multiplies a  dimension-6 coefficient, since the difference is a dimension-8 effect.  
On the other hand, when $v_T$ appears in a dimension-4 term, it must be replaced by (\ref{eq:vevrep}).
Finally, the quantity $\lambda$ in the Higgs potential can be eliminated
in terms of the input parameters (\ref{eq:InputPar}) according to
\begin{align}
\lambda = \frac{m_H^2}{2\hat{v}_T^2}
\left[1 -2 C_{H,\rm kin}
+ 2\hat{v}_T^2 \frac{\hat{c}_w}{\hat{s}_w}\left(C_{HWB}+\frac{\hat{c}_w}{4\hat{s}_w}C_{HD}\right)
+\frac{3\hat{v}_T^4}{m_H^2}C_H  \right] \,.
\end{align}

\subsection{Gauge fields}
\label{sec:GaugeNormalization}
In the following section we review the rotation to the mass basis of the gauge fields in SMEFT, closely following the procedure in \cite{Alonso:2013hga}. We denote the covariant derivative in the electroweak sector of the SM by
\begin{equation}
\label{eq:SMCoD}
D_\mu = \partial_\mu - i (g \tau)^a A^a_\mu \, ,
\end{equation}
where $A^a_\mu=(W^1_\mu,W^2_\mu,W^3_\mu,B_\mu)$, and the generators are denoted $(g \tau)^a=(g_2 \tau^1, g_2 \tau^2, g_2 \tau^3, g_1 Y)$, where $\tau^I = \sigma^I/2$ with $\sigma^I$ the Pauli matrices and $Y$ the hypercharge.

When including dimension-6 operators we must first redefine the gauge fields as
\begin{equation}
\label{eq:GaugeRedefine}
B_{\mu}=\left(1+\hat{v}_T^2 C_{HB}\right)\mathcal{B}_\mu \, , \qquad W^I_\mu = \left(1+\hat{v}_T^2 C_{HW}\right)\mathcal{ W}^I_{\mu} \, ,
\end{equation}
to ensure correct gauge field normalization. Additionally, we modify the couplings as
\begin{equation}
\label{eq:ModifiedCouplings}
\bar{g}_1=(1+\hat{v}_T^2C_{HB}) g_1 \, , \qquad \bar{g}_2= (1+\hat{v}_T^2C_{HW}) g_2 \, ,
\end{equation}
such that the combinations $g_1 B_\mu=\bar{g}_1\mathcal{B}_\mu $ and 
$g_2 W^I_\mu=\bar{g}_2\mathcal{W}^I_\mu $ remain unchanged. It can be shown that $\bar{g}_1$ and $\bar{g}_2$ can be written in terms of the physical input parameters listed in (\ref{eq:InputPar}) as
\begin{align}
\bar{g}_1= \frac{e}{\hat{c}_w}\left(1- \frac{\hat{v}_T^2}{4}C_{HD}\right)\,
 , \qquad \bar{g}_2=\frac{e}{\hat{s}_w}\left(1 + \hat{v}_T^2 \frac{\hat{c}_w}{\hat{s}_w}\left[C_{HWB}+\frac{\hat{c}_w}{4\hat{s}_w}C_{HD} \right]\right)  \, .
\end{align}

The class-4 operator $Q_{HWB}$ introduces a kinetic mixing term between the $\mathcal{W}_\mu^3$ and $\mathcal{B_\mu}$ gauge fields not seen in the SM, which is of the form $\sim - \frac{1}{2}v_T^2 \mathcal{W}^3_\mu \mathcal{B}_\mu $. This term can be removed by a linear shift in these fields, which proceeds as
\begin{align}
\label{eq:GaugeShift}
\mathcal{A}^a_\mu= M^{ab}A'^b_\mu \, ,
\end{align}
where $\mathcal{A}^a_\mu=(\mathcal{W}^1_\mu,\mathcal{W}^2_\mu,\mathcal{W}^3_\mu,\mathcal{B}_\mu)$, $A'^a_\mu=(W'^1_\mu,W'^2_\mu,W'^3_\mu,B'_\mu)$ and 
\begin{align}
\label{eq:MMatrix}
M=
\begin{pmatrix}
\mathbf{1}_{2\times 2} & \mathbf{0}_{2\times 2} \\
\mathbf{0}_{2\times 2} & m
\end{pmatrix} \, , 
\qquad
m= \begin{pmatrix}
1 & - \frac{1}{2}v_T^2 C_{HWB} \\
-\frac{1}{2}v_T^2 C_{HWB} & 1 
\end{pmatrix} \, ,
\end{align}
such that the new `primed' gauge fields have diagonal kinetic terms.
 These are rotated to the mass basis according to 
 \begin{align}
A'^a_\mu=R^{ab} \tilde{A}_\mu^b \, ,
\end{align}
where $\tilde{A}_\mu^a$ comprises the physical gauge fields as $\tilde{A}_\mu=(\mathcal{W}^+_\mu,\mathcal{W}^-_\mu, \mathcal{Z}_\mu, \mathcal{A}_\mu)$, and $R$ is given by
\begin{align}
\label{eq:RotationMatrix}
R=
\begin{pmatrix}
\frac{1}{\sqrt{2}} & \frac{1}{\sqrt{2}} & 0 & 0 \\
\frac{i}{\sqrt{2}} & -\frac{i}{\sqrt{2}} & 0 & 0 \\
0 & 0 & \overline{c}_w & \overline{s}_w \\
0 & 0 & -\overline{s}_w & \overline{c}_w
\end{pmatrix} \, ,
\end{align}
\begin{align}
\bar{c}_w = \hat{c}_w\left(1+ \frac{\hat{v}_T^2}{4}C_{HD}+   \frac{\hat{s}_w\hat{v}_T^2}{2\hat{c_w}}C_{HWB}\right) \, ,
\qquad \bar{s}_w^2 = 1 -  \bar{c}_w^2  \,.
\end{align}
With this notation, the relation between the weak-basis fields $\mathcal{A}^a_\mu$ and 
the mass basis fields $\tilde{A}^a_\mu$ is
\begin{align}
\label{eq:GaugeWeak-Mass}
\mathcal{A}^a_\mu=M^{ab} R^{bc} \tilde{A}^c_\mu \, .
\end{align}
In terms of the input parameters in (\ref{eq:InputPar}), the explicit definitions of the
photon and $Z$-boson fields in terms of the weak-basis fields is
\begin{align}
\begin{pmatrix}
\mathcal{W}^3_\mu \\
\mathcal{B}_\mu \end{pmatrix}
=
\begin{pmatrix}
\hat{c}_w+\frac{1}{4}\hat{c}_w \hat{v}_T^2 \left(C_{HD}+4 \frac{\hat{s}_w}{\hat{c}_w}C_{HWB} \right) \qquad \quad & \hat{s}_w-\frac{\hat{c}_w^2 \hat{v}_T^2}{4\hat{s}_w} \left(C_{HD}+4 \frac{\hat{s}_w}{\hat{c}_w}C_{HWB} \right)\\
-\hat{s}_w+\frac{\hat{c}_w^2 \hat{v}_T^2}{4 \hat{s}_w}C_{HD} & \hat{c}_w+\frac{\hat{c}_w\hat{v}_T^2}{4}C_{HD}
\end{pmatrix}
\begin{pmatrix}
\mathcal{Z}_\mu \\
\mathcal{A}_\mu 
\end{pmatrix} 
\, .
\end{align}
Furthermore,  the dimension-6 SMEFT covariant derivative in the mass basis is given by 
\begin{align}
D_\mu = \partial_\mu -& i \frac{e}{\hat{s}_w} \left[ 1 + \frac{\hat{c}_w^2 \hat{v}_T^2}{4 \hat{s}_w^2} C_{HD} + \frac{\hat{c}_w \hat{v}_T^2}{\hat{s}_w} C_{HWB} \right] \left( \mathcal{W}_\mu ^+ \tau^+ + \mathcal{W}_\mu ^- \tau^- \right) \notag \\
	-& i \left[ \frac{e}{\hat{c}_w \hat{s}_w}\left(1 + \frac{(2 \hat{c}_w^2-1) \hat{v}_T^2}{4 \hat{s}_w^2}C_{HD} + \frac{\hat{c}_w \hat{v}_T^2}{\hat{s}_w}C_{HWB} \right)\left(\tau^3- \hat{s}_w^2 Q \right) \right. \notag \\
		+& \left. e \left(\frac{\hat{c}_w \hat{v}_T^2}{2 \hat{s}_w}C_{HD}+ \hat{v}_T^2 C_{HWB} \right)Q  \right] \mathcal{Z}_\mu -i e Q \mathcal{A}_\mu \, ,
\end{align}
where $Q=\tau^3+Y$ and $\tau^\pm=(\tau^1 \pm i \tau^2)/\sqrt{2}$.

\subsection{Gauge fixing in $R_\xi$ gauges}
\label{sec:GaugeFixing}
Gauge fixing in SMEFT has been discussed in~\cite{Dedes:2017zog,Helset:2018fgq,Misiak:2018gvl}.
In this section we explain our own implementation, which we have used 
when verifying the gauge independence of the decay rate and counterterms with explicit one-loop 
computations.  
Throughout this section we follow closely the notation used for gauge fixing in the SM as presented 
in \cite{Peskin:1995ev}.    We parametrise the Higgs doublet in terms of real scalar fields as
\begin{align}
\label{eq:SMDoublet}
H= \frac{1}{\sqrt{2}}
\begin{pmatrix}
-i(\phi_1-i \phi_2) \\
\phi_4+i \phi_3
\end{pmatrix} \, ,
\end{align}
and use the real representation of the generators, ${T^a=-i \tau^a}$, where the $\tau^a$ were defined 
below (\ref{eq:SMCoD}).  We expand each $\phi_i$ about its vacuum expectation value, denoted $\langle \phi_i\rangle=\phi_{0_i}$ as 
\begin{align}
\label{eq:phiExpansion}
\phi_i=\phi_{0_i}+\chi_i \, ,
\end{align}
where  $\chi_{i\neq 4}$ are the Goldstone bosons, $\chi_4$ is related to the physical Higgs boson, $h$,
and $\phi_{0_i}= \delta_{i4} v_T/\sqrt{2}=(0,1)^Tv_T/\sqrt{2}$.
In $R_\xi$ gauges one aims to remove the Goldstone-gauge boson mixing terms, which in the SM 
take the form
\begin{align}
\label{eq:SMGaugeGoldstone}
\mathcal{L} \supset (\partial^\mu \chi_i)A^a_\mu (g T)^a_{ij} \phi_{0_j} \, ,
\end{align} 
where $(gT)^a=(g_2 T^1,g_2 T^2,g_2 T^3,g_1 T^4)$. The $i=4$ component in (\ref{eq:SMGaugeGoldstone}) gives no contribution to the Lagrangian. 

We now include dimension-6 effects in SMEFT.  We begin by defining the canonically-normalized fields of the Higgs doublet in (\ref{eq:HiggsDoublet}) in terms of those in (\ref{eq:phiExpansion}) via the transformation
\begin{align}
\label{eq:chiShift}
\chi_i=X_{ij} \chi'_j \, , \qquad 
X=\begin{pmatrix}
1 & 0 & 0 & 0 \\
0 & 1 & 0 & 0 \\
0 & 0 & 1-\frac{1}{4}\hat{v}_T^2 C_{HD} & 0 \\
0 & 0 & 0 & 1+C_{H, \text{kin}} 
\end{pmatrix} \, ,
\end{align}
such that the $\chi'_i$ are related to the fields in (\ref{eq:HiggsDoublet}) by 
\begin{align}
\chi'_1= \frac{1}{\sqrt{2}}(\phi^+ +\phi^-) \, ,\quad \chi'_2=\frac{i}{\sqrt{2}}(\phi^+ -\phi^-) \, ,\quad \chi'_3=\phi^0 \, , \quad \chi'_4=h \, .
\end{align}
Moreover, we replace the gauge fields and couplings as in (\ref{eq:GaugeRedefine}), (\ref{eq:ModifiedCouplings}) and (\ref{eq:GaugeShift}) such that all the Goldstone-gauge mixing terms of the SMEFT Lagrangian may be written
\begin{align}
\label{eq:GaugeGoldMixingSMEFT}
\mathcal{L} &\supset (X_{ik}\partial^\mu \chi'_k) A'^a_\mu (\overline{g} T')^a_{ij} \phi_{0_j}+\frac{1}{2}v_T^2 C_{HD}(\partial^\mu \chi'_3) A'^a_\mu (\overline{g} T')^a_{3j} \phi_{0_j} \notag \\
 &=(\partial^\mu \chi'_i) A'^a_\mu (\overline{g} \mathcal{F})^a_{\, i} \, ,
\end{align}
where the second term on the first line of (\ref{eq:GaugeGoldMixingSMEFT}) is the contribution arising from the explicit presence of the $C_{HD}Q_{HD}$ term in the dimension-6 SMEFT Lagrangian.
Here we have introduced the object $(\overline{g}T)^a$, which is defined similarly to $(gT)^a$ in (\ref{eq:SMGaugeGoldstone}), but with all instances of the gauge couplings replaced as $g_i \rightarrow \overline{g}_i$, and further defined `primed' generators 
\begin{align}
\label{eq:PrimedGenerators}
    (\overline{g}T')^a &= M^{ab}(\overline{g}T)^b \notag \\
    &= \left(\overline{g}_2 T^1, \, \overline{g}_2 T^2, \, \overline{g}_2 T^3- \frac{1}{2} \overline{g}_1 v_T^2 C_{HWB} T^4, \, \overline{g}_1 T^4- \frac{1}{2} \overline{g}_2 v_T^2 C_{HWB} T^3 \right) \, ,
\end{align}
where $M^{ab}$ is given in (\ref{eq:MMatrix}), and also the object 
\begin{align}
(\overline{g} \mathcal{F})^a_{\, i} &=
X_{ij}(\overline{g} T')^a_{jk} \phi_{0_k} +\delta_{i3} \frac{v_T^2}{2}C_{HD} (\overline{g}T')^a_{3k}\phi_{0_k} \nonumber \\
&=(X^{-1})_{ij} (\overline{g}T')^a_{jk} \phi_{0_k} \, ,
\end{align}
where in the final line we have used that $X$ has only diagonal elements, $X_{11}=X_{22}=(X^{-1})_{11}=(X^{-1})_{22}=1$, $(1+\frac{\hat{v}_T^2}{2} C_{HD})X_{33}=(X^{-1})_{33}$ and that the $X_{44}$ component gives no contribution. 
In order to calculate the matrix $(\overline{g}\mathcal{F})^a_{\, i}$ we use, for example, that $(\overline{g}T')^1 \phi_0$ equals $\overline{g}_2 v_T/2$ times a unit vector in the $\phi^1$ direction.  One finds
\begin{align}
\label{eq:SMEFTgF}
(\overline{g}\mathcal{F})^a_{\, i}=
\frac{v_T}{2}\begin{pmatrix}
\overline{g}_2 & 0 & 0 & 0 \\
0 & \overline{g}_2 & 0 & 0 \\
0 & 0 &  \overline{g}_2(1+ \frac{\hat{v}_T^2}{4}C_{HD}) + \overline{g}_1 \frac{\hat{v}_T^2}{2} C_{HWB} & 0 \\
0 & 0 & - \overline{g}_1(1+ \frac{\hat{v}_T^2}{4}C_{HD})-\overline{g}_2 \frac{\hat{v}_T^2}{2} C_{HWB} & 0
\end{pmatrix} \, .
\end{align}

We follow the Faddeev-Popov gauge-fixing procedure such that the SMEFT gauge-fixed generating functional $Z$ takes the form
\begin{align}
\label{eq:GaugeFixingFunctional}
Z=C \int \mathcal{D} A' \mathcal{D} \chi' \exp \left[ i \int d^4x \left( \mathcal{L} \left[A',\chi' \right] -\frac{1}{2}(G)^2 \right) \right] \text{det} \left(\frac{\delta G}{\delta (\alpha'/\overline{g})} \right) \, ,
\end{align}
where $G^a$ is the gauge-fixing function and the object $(\alpha'/\overline{g})^b$ is defined below. We choose the gauge-fixing function in (\ref{eq:GaugeFixingFunctional}) as
\begin{align}
\label{eq:GaugeFixingF}
G^a = \frac{1}{\sqrt{\xi}} \left( \partial^\mu A'^a_\mu- \xi (\overline{g}\mathcal{F})^a_{\, i} \chi'_i  \right) \, ,
\end{align}
which defines the $R_\xi$ gauges in SMEFT.\footnote{ Note that in principle we can have a different $\xi$ for each of the physical gauge fields.} We see that the form of the gauge-fixing function in (\ref{eq:GaugeFixingF}) resembles that of the $R_\xi$ gauges in the SM with the gauge fields replaced by their primed counterparts and $F$ replaced with $\mathcal{F} \,$. The Goldstone-gauge boson mixing terms in (\ref{eq:GaugeGoldMixingSMEFT}) are then removed by the $-\frac{1}{2}(G)^2$ term in (\ref{eq:GaugeFixingFunctional}).

Interactions of SM particles with ghost fields arise through the functional determinant in (\ref{eq:GaugeFixingFunctional}), for which we must determine the variation of $G^a$ under arbitrary gauge transformations. The gauge transformation of the scalar fields may be written
\begin{align}
\label{eq:ScalarGaugeTransform}
\delta \phi_i =- \alpha^a T^a_{ij} \phi_j \equiv - \left(\frac{\alpha}{\overline{g}} \right)^a (\overline{g} T)^a_{ij} \phi_j \equiv -\left(\frac{\alpha'}{\overline{g}} \right)^a (\overline{g} T')^a_{ij} \phi_j \, ,
\end{align}
where the second relation defines the object $(\alpha/\overline{g})^a$ and the third relation defines the object $(\alpha'/\overline{g})^a$ as
\begin{align}
\label{eq:aMap}
\left(\frac{\alpha}{\overline{g}}\right)^a= M^{ab} \left(\frac{\alpha'}{\overline{g}}\right)^b \, .
\end{align}
We may use (\ref{eq:chiShift}) and (\ref{eq:ScalarGaugeTransform}) to find the gauge transformation of $\chi'_i$:
\begin{align}
\label{eq:chiGaugeTransformation}
\delta \chi'_i=(X^{-1})_{ij} \delta \chi_j &= -\left( \frac{\alpha'}{\overline{g}} \right)^a (X^{-1})_{ij} (\overline{g} T')^a_{jk} ( \phi_{0_k}+X_{kl} \chi'_l) \notag \\
	&\equiv -\left( \frac{\alpha'}{\overline{g}} \right)^a \left( (\overline{g} \mathcal{F})^a_{\, i} + (\overline{g} \mathcal{T})^a_{ij} \chi'_j \right) \, ,
\end{align}
where we have defined the object $(\overline{g}\mathcal{T})^a_{ij} \equiv (X^{-1})_{ik}(\overline{g}T')^a_{kl} X_{lj}$ . Explicitly $(\overline{g}\mathcal{T})^a_{ij}$ acts on $\chi'_i$ as (for brevity and as no other terms enter our calculation, we give only the Higgs contributions to this term)
\begin{align}
(\overline{g} \mathcal{T})^a_{ij}\chi'_j \supset
\frac{h}{2}\begin{pmatrix}
\overline{g}_2(1+ C_{H,\text{kin.}}) & 0 & 0 & 0 \\
0 & \overline{g}_2(1+C_{H,\text{kin.}}) & 0 & 0 \\
0 & 0 & \overline{g}_2(1+\hat{v}_T^2 C_{H \Box})+\overline{g}_1 \frac{\hat{v}_T^2}{2} C_{HWB} & 0 \\
0 & 0 & -\overline{g}_1(1+\hat{v}_T^2 C_{H \Box})-\overline{g}_2 \frac{\hat{v}_T^2}{2} C_{HWB} & 0
\end{pmatrix} \, .
\end{align}
We may similarly write the transformation of the unprimed gauge fields as
\begin{align}
\label{eq:GaugeGaugeTransformation}
\delta A^a_\mu = \partial_\mu  \left(\frac{\alpha}{\overline{g}} \right)^a-f^{abc} \alpha^b A_\mu^c &\equiv \partial_\mu  \left(\frac{\alpha}{\overline{g}} \right)^a- \overline{g}_2 f^{abc} \left(\frac{\alpha}{\overline{g}} \right)^b A_\mu^c  \, .
\end{align}
The object $f^{abc}= \epsilon^{abc}$ if $a,b,c\in 1,2,3$ and vanishes otherwise, which we have used to replace
$\alpha^b \to  \overline{g}_2(\alpha/\overline{g})^b$ in the above equation.  
The form of $\delta A'^a_\mu$ in terms of the object $(\alpha'/\overline{g})^a$ is then found using (\ref{eq:GaugeShift}), (\ref{eq:aMap}) and (\ref{eq:GaugeGaugeTransformation})
\begin{align}
\label{eq:GaugeGaugePrimedTransform}
\delta A'^a_\mu = (M^{-1})^{ab} \delta A^b_\mu = \partial_\mu \left( \frac{\alpha'}{\overline{g}} \right)^a-\overline{g}_2 (M^{-1})^{ab} f^{bcd} M^{cc'} \left( \frac{\alpha'}{\overline{g}} \right)^{c'} A_\mu^d \, .
\end{align}
We can now calculate the functional derivatives needed to evaluate (\ref{eq:GaugeFixingFunctional}) using the results in (\ref{eq:chiGaugeTransformation}) and (\ref{eq:GaugeGaugePrimedTransform}). First, one has 
\begin{align}
\label{eq:APrimedVar}
\frac{\delta A'^a_\mu}{\delta (\alpha'/\overline{g})^b} \equiv \mathcal{M}^{ab}_\mu= \delta^{ab} \partial_\mu -\overline{g}_2 (M^{-1})^{a b'} f^{b' c d} A_\mu^d M^{cb} \, ,
\end{align}
(note that the gauge fields here are the unprimed gauge fields), where the explicit result is
\begin{align}
\mathcal{M}^{ab}_\mu = \overline{g}_2
\begin{pmatrix}
\frac{1}{\bar{g}_2}\partial_\mu & W^3_\mu & -W^2_\mu \, \, \, \, \,  & \frac{1}{2}\hat{v}_T^2 C_{HWB} W^2_\mu \\
-W^3_\mu & \frac{1}{\bar{g}_2}\partial_\mu & W^1_\mu  & -\frac{1}{2}\hat{v}_T^2 C_{HWB} W^1_\mu \\
W^2_\mu & -W^1_\mu & \frac{1}{\bar{g}_2}\partial_\mu & 0 \\
\frac{1}{2}\hat{v}_T^2 C_{HWB} W^2_\mu \, \, \, \, \,  & -\frac{1}{2}\hat{v}_T^2 C_{HWB} W^1_\mu & 0 & \frac{1}{\bar{g}_2}\partial_\mu 
\end{pmatrix} \, .
\end{align}
From (\ref{eq:APrimedVar}) and (\ref{eq:chiGaugeTransformation}), the variation of the gauge-fixing function, $G^a$ in (\ref{eq:GaugeFixingF}) is
\begin{align}
\label{eq:GVar}
\frac{\delta G^a}{\delta (\alpha'/\overline{g})^b}= \frac{1}{\sqrt{\xi }}  \left( \partial^\mu \mathcal{M}^{ab}_\mu + \xi (\overline{g}\mathcal{F})^a_{\, i}  \left( (\overline{g}\mathcal{F})^b_{\, i}+(\overline{g} \mathcal{T})^b_{ij} \chi'_j \right) \right) \,.
\end{align}
Following the usual procedure the ghost Lagrangian is 
\begin{align}
\label{eq:GhostL}
\mathcal{L}_\text{ghost} = \overline{c}^a \left[-\left(\partial^\mu \mathcal{M}^{ab}_\mu\right) - \xi (\overline{g}\mathcal{F})^a_{\, i}  \left( (\overline{g}\mathcal{F})^b_{\, i}+(\overline{g} \mathcal{T})^b_{ij} \chi'_j \right) \right] c^b   \, .
\end{align}
The ghost fields in (\ref{eq:GhostL}) are given by $c^a=(c_{W^1},c_{W^2},c_{W^3},c_{B})$, and similarly for the fields in $\overline{c}^a$. The form of the ghost mass matrix in (\ref{eq:GhostL}) is
\begin{align}
(m^2_\text{ghost})^{ab}=\xi (\overline{g}\mathcal{F})^a_{\, i} (\overline{g}\mathcal{F})^b_{\, i} \, ,
\end{align}
which is diagonalized by the matrix $R$ in (\ref{eq:RotationMatrix}) such that
\begin{align}
(m^2_{D,\text{ghost}})^{ab} \equiv (R^{-1})^{ac}(m^2_\text{ghost})^{cd}R^{db}=\text{diag}(M_W,M_W,M_Z,0) \, .
\end{align}
The ghosts in the mass basis, denoted $u^a$ and $\overline{u}^a$, are thus related to those in the weak basis by
\begin{align}
c^a=R^{ab}u^b \, , \qquad \overline{c}^a=\overline{u}^b (R^{-1})^{ba} \, ,
\end{align}
where $u^a=(u_{W^+},u_{W^-},u_Z,u_A)$, and similarly for $\overline{u}^a$. With the gauge fields $A_\mu$ written in terms of the mass basis as described in (\ref{eq:GaugeWeak-Mass}), the ghost Lagrangian in the mass basis is therefore
\begin{align}
\label{eq:GhostLMass}
\mathcal{L}_\text{ghost}=\overline{u}^a \left[-\left((R^{-1})^{ac} \partial^\mu \mathcal {M}^{cd}_\mu R^{db}\right) - \xi \left( (m^2_{D,\text{ghost}})^{ab}+(R^{-1})^{ac} (\overline{g}\mathcal{F})^c_{\, i}(\overline{g} \mathcal{T})^d_{ij} \chi'_j R^{db} \right) \right] u^b   \, .
\end{align}
Although our derivation is rather different, we find that the Feynman rules produced by the Lagrangian in (\ref{eq:GhostLMass}) exactly match those found in \cite{Dedes:2017zog}.

\subsection{Yukawa sector}
\label{sec:Yuk_app}
The fermion masses in SMEFT involve the Wilson coefficients of class-5 operators as well as the SM Yukawa
matrices.    The relevant part of the Lagrangian (following the convention used in~\cite{Jenkins:2013zja,Jenkins:2013wua,Alonso:2013hga}) is given by
\begin{align}
\mathcal{L} \supset& -\left[ [Y_u]_{r_1 r_2} \tilde{H}^{\dagger j} \overline u_{r_1} \,   q_{r_2 j} + [Y_d]_{r_1 r_2} H^{\dagger j} \overline d_{r_1} \,   q_{r_2 j} + [Y_e]_{r_1 r_2} H^{\dagger j} \overline e_{r_1} \,   l_{r_2 j}+ \text{h.c.} \right] 
\nonumber \\[1mm] &
+\left[ C^*_{\substack{u H \\ r_2 r_1}} (H^\dagger H)  \tilde{H}^{\dagger j} \overline u_{r_1} \,   q_{r_2 j} + C^*_{\substack{d H \\ r_2 r_1}} (H^\dagger H) H^{\dagger j} \overline d_{r_1} \, q_{r_2 j} + C^*_{\substack{e H \\ r_2 r_1}} (H^\dagger H) H^{\dagger j} \overline e_{r_1} \, l_{r_2 j} +\text{h.c.}\right], \label{eq:smeft_fullYuk}
\end{align}
where the subscripts $j$ and $r_i$ are $SU(2)$ and generation indices respectively.
In what follows we perform rotation to the mass basis using the down-type quarks as an example and
suppress the explicit addition of the hermitian conjugate ($\text{+h.c.}$). 
After spontaneous symmetry breaking in unitary gauge and keeping only dimension-6 terms
one finds
\begin{align}
\mathcal{L}_{\text{mass}} &= -\frac{v_T}{\sqrt{2}} \overline{d}_{R r_1} \bigg(\left[Y_d\right]_{r_1 r_2}-\frac{v_T^2}{2} C^*_{\substack{d H \\ r_2 r_1}} \bigg) d_{L r_2} \equiv -\overline{d}_{R r_1} \left[M_d\right]_{r_1 r_2}d_{L r_2} \, , \label{eq:smeft_mass} \\
\mathcal{L}_{\text{yuk}} &= -\frac{1}{\sqrt{2}} h \,  \overline{d}_{R r_1}\bigg(\left[Y_d\right]_{r_1 r_2}\left[1+C_{H,\text{kin}}\right] -\frac{3}{2}v_T^2 C^*_{\substack{d H \\ r_2 r_1}} \bigg)d_{L r_2} \, ,\label{eq:smeft_yuk}
\end{align}
where $\mathcal{L}_{\text{yuk}}$ is defined as the term proportional to the $h \overline{\psi} \psi$ operator. Additionally we have included the subscripts $L$ and $R$ on the quark fields to denote their handedness.
As usual, we perform rotations on the quark fields to go to the mass basis
\begin{equation}
d_{R r_1} \rightarrow \left[U_{d_R}\right]_{r_1 r_2} d_{R r_2} \, , \qquad d_{L r_1} \rightarrow \left[U_{d_L}\right]_{r_1 r_2} d_{L r_2} \, ,
\end{equation}
such that
\begin{equation}
\left[U^\dagger_{d_R} M_d U_{d_L}\right]_{r_1 r_2}=\left[m_d\right]_{r_1 r_2} \, ,
\end{equation}
where $\left[m_d\right]=\text{diag}(m_d,m_s,m_b)$. 
After the field rotation, the $h\bar{\psi}\psi$ term becomes
\begin{equation}
\mathcal{L}_{\text{yuk}} = -\frac{1}{\sqrt{2}} h \,  \overline{d}_{R r_1}\bigg(\left[1+C_{H,\text{kin}}\right]
\frac{\sqrt{2}}{v_T}\left[m_d\right]_{r_1 r_2} -v_T^2 \left[C_{dH}^{m\dagger}\right]_{r_1 r_2} \bigg)d_{L r_2} \, ,
\label{eq:smef_yukmass_nonMFV}
\end{equation}
where
\begin{equation}
\left[C_{dH}^{m\dagger}\right]_{r_1 r_2} = \left[U_{d_R}^\dagger C^\dagger_{d H} U_{d_L}\right]_{r_1 r_2} \, .
\label{eq:cdh_massbasis}
\end{equation}
Thus, in the mass basis the Wilson coefficients contributing to $h\bar{\psi}\psi$ couplings are a linear combination of those in the weak eigenstate basis.  Similar results can be derived for any Wilson coefficient 
$C^m_i$ multiplying a mass-basis operator containing fermions.

Note that in contrast to the SM, SMEFT contains flavour-violating Higgs couplings even in the mass basis.  However, in our calculation we approximate the CKM matrix by the unit matrix,  in which case these flavour-violating couplings  do not contribute to the NLO $h\to b\bar{b}$ decay rate at dimension-6.  In fact, within this approximation there 
are no transitions between fermion generations, which allows us to introduce a compact notation for 
Wilson coefficients such as (\ref{eq:cdh_massbasis}) which multiply mass-basis operators.  First, for operators
involving  right-handed fields we can always indicate the generation by the explicit flavour.  Examples of this 
are
\begin{equation}
C_{bH} \equiv C_{\substack{dH \\ 33}}^{m} \, ,  \quad C_{H\mu} \equiv C^{m}_{\substack{He \\ 22} } \, ,
\quad C_{tW} \equiv C^{m}_{\substack{uW \\ 33}} \, ,
\end{equation}
and similarly for any fermion $f$. Some Wilson coefficients for operators containing left-handed fields use the subscripts $q_r$ and $\ell_r$, so it is not possible to indicate the doublet generation $r$ through the flavours
it contains.
However, the third generation plays a prominent role in our calculation, so our convention
is to suppress any dependence on $r=3$ but display explicitly the flavour indices only on operators involving
first- and second-generation fermions, which appear through electroweak boson self-energies and tadpoles.  
Examples of operators in this notation are 
\begin{align}
C_{qtqb}^{(1)} \equiv C^{m(1)}_{\substack{quqd \\ 3333}}  \,,  \quad
\left[C_{Hq}^{(1)}\right]_{22}   \equiv \left[C^{m(1)}_{Hq}\right]_{22} \, ,
\end{align}
where the first coefficient multiplies a mass-basis operator with field content $\bar{t}t\bar{b}b$ and the second
coefficient multiplies a mass-basis operators with fermion content $c\bar{c}$ and $s\bar{s}$.  

An important feature of SMEFT in the mass basis is that couplings between left and right-handed fields are 
not always associated with powers of the fermion mass, as in the SM.  For instance, the $C_{bH}$ operator contains 
a $hbb$ coupling which is not proportional to the $b$-quark Yukawa, which is $y_b \approx \sqrt{2}m_b/\hat{v}_T$ in
the mass basis.  For this reason $h\to b\bar{b}$ offers an important probe on the flavour structure of SMEFT.
However, in this work we are interested in the structure of NLO contributions in SMEFT rather than questions
of flavour, so in our numerical analysis it is convenient to display results in such a way that all contributions to 
the decay rate multiply a symbolic factor of $m_b^2/\hat{v}_T^2$ as in the SM.  We emphasize that this is not 
a restriction of our calculation but rather a matter of convenience.  However, if the Wilson coefficients are generated
by a new physics scenario which respects  Minimal Flavour Violation (MFV)~\cite{DAmbrosio:2002vsn} it is 
something which occurs naturally. See refs~\cite{Alonso:2013hga,Brivio:2017btx} for further discussion on this in the context of SMEFT.

\section{Analytic results}
\label{sec:analytic}
In this section we give analytic results for the LO decay rate and the NLO QCD-QED 
corrections $\Gamma_{g,\gamma}$ in the small-$m_b$ limit  used in our numerical analysis, as well as the large-$m_t$ corrections $\Gamma_{t}$.  
We give results which can be easily converted 
between the on-shell and \msbar~schemes for $X\in\{m_b,e\}$ using the notation in (\ref{eq:CXdef}).
We will also need  to split the finite part of the counterterms in the on-shell scheme 
into QCD-QED, large-$m_t$, and remaining pieces. To do so we define
\begin{align}
\label{eq:BEdef}
   \frac{\delta m_b^{(i)\rm O.S., fin.}}{m_b} &= \delta b^{(i)}_{g,\gamma}+\delta b^{(i)}_t+\delta b^{(i)}_{\rm rem}  \,, \nonumber \\
\frac{\delta e^{(i)\rm O.S., fin.}}{e} &= \delta e^{(i)}_{g,\gamma}+\delta e^{(i)}_t+\delta e^{(i)}_{\rm rem} \, ,
\end{align}
where the superscript $i=4,6$ labels the NLO contribution from dimension-$i$ operators.  We use 
this notation throughout the section.

\subsection{LO decay rate}

The LO contributions to the decay rate as defined in  (\ref{eq:PertExpansion}) 
are given by  
\begin{align}
\label{eq:LOgam}
 \Gamma^{(4,0)}& = \frac{N_c m_H m_b^2}{8 \pi \hat{v}_T^2} \, ,\\
\Gamma^{(6,0)}& = 2 \Gamma^{(4,0)}
\left[C_{H\Box}-\frac{C_{HD}}{4}\left(1- \frac{\hat{c}_w^2}{\hat{s}_w^2} \right)
+\frac{\hat{c}_w}{\hat{s}_w} C_{HWB}  
-\frac{\hat{v}_T}{m_b} \frac{C_{bH}}{\sqrt{2}}\right]\hat{v}_T^2 \,.
\end{align}

\subsection{QCD-QED corrections}
\label{sec:Gqcd-qed}
 
\begin{figure}[t]
\centering
\begin{multicols}{3}

\includegraphics[scale=1]{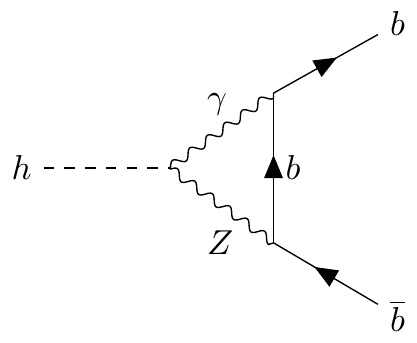} \\
\includegraphics[scale=1]{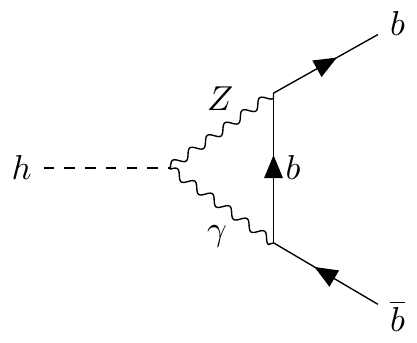} \\ 
\includegraphics[scale=1]{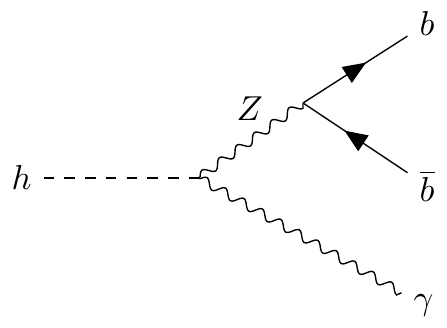} 
\end{multicols}
\vspace{-7mm}
\begin{multicols}{3}
(a) \\
(b) \\
(c)
\end{multicols}
\caption{Real (a, b) and virtual (c) corrections to the $h \rightarrow b \overline{b}$ decay rate due to the $h\gamma Z$ vertex generated by the operators $Q_{HB}$, $Q_{HW}$ and $Q_{HWB}$. }
\label{fig:RealEmissions}
\end{figure}

The NLO result for the QCD-QED corrections in the SM can be written as 
\begin{align}
 \Gamma_{g,\gamma}^{(4,1)} = \overline{\Gamma}_{g,\gamma}^{(4,1)}
 +2 c_{m_b} \Gamma^{(4,0)} \delta b^{(4)}_{g,\gamma} \, ,
 \end{align}
while that in SMEFT takes the form
\begin{align}
\Gamma_{g,\gamma}^{(6,1)}&  =\overline{\Gamma}_{g,\gamma}^{(6,1)}   + 2 c_{m_b} \Gamma^{(4,0)}
 \delta b_{g,\gamma}^{(4)}\left( \frac{C_{bH}\hat{v}_T^2}{\sqrt{2}}\frac{\hat{v}_T}{m_b}+ \frac{\Gamma^{(6,0)}}{\Gamma^{(4,0)}} \right) \,,
  \end{align}
where 
\begin{align}
\delta b_{g,\gamma}^{(4)}
=-\left(\frac{C_F \alpha_s+ Q_b^2 \alpha}{\pi}\right)
\left[1+\frac{3}{4}\ln\left(\frac{\mu^2}{m_b^2}\right)\right] \, ,
\end{align}
and we have used that $\delta b_{g,\gamma}^{(6)}=0$ in the small-$m_b$ limit.

The $\overline{\Gamma}_{g,\gamma}$ are the QCD-QED corrections to the decay rates in the \msbar~scheme for 
$m_b$.  The QCD corrections were obtained in \cite{Gauld:2016kuu}. Most of the QED corrections  can be derived from those results by making appropriate replacements. 
The exception is the contribution  proportional to the $h\gamma Z$ vertex in SMEFT, which arises
from the real and virtual emission diagrams in Figure~\ref{fig:RealEmissions} and has no analogue in QCD.  
We have obtained the contributions from these diagrams to the decay rate by evaluating and adding together the  virtual and real corrections as in (\ref{eq:RealPlusVirtual}).  This new result together with the other
QCD-QED corrections in the small-$m_b$ limit can be written as
\begin{align}
\label{eq:QEDQCDGam61}
 \overline{\Gamma}_{g,\gamma}^{(4,1)} & = \Gamma^{(4,0)}\left(\frac{C_F \alpha_s+ Q_b^2 \alpha}{\pi}\right) \left[\frac{17}{4}+\frac{3}{2}\ln\left(\frac{\mu^2}{m_H^2}\right)\right] \, , \nonumber \\
\overline{\Gamma}_{g,\gamma}^{(6,1)} & = 
\Gamma^{(6,0)}  \frac{\overline{\Gamma}_{g,\gamma}^{(4,1)}}{\Gamma^{(4,0)}} + 
 \frac{\hat{v}_T^2 }{\pi}\Gamma^{(4,0)}\bigg\{  \frac{m_H^2}{\sqrt{2}\hat{v_T} m_b}
  \bigg(\frac{C_F}{g_s} \alpha_s C_{bG}+ \frac{Q_b}{\overline{e}^{(\ell)}} \alpha
  \left( C_{bB} \hat{c}_w - C_{bW} \hat{s}_w\right)\bigg) 
\nonumber \\ &
+\left(C_F \alpha_s C_{HG}+ Q_b^2 \alpha\, c_{h\gamma\gamma}\right) 
  \left[19-\pi^2 + \ln^2\left(\frac{m_b^2}{m_H^2}\right)+
  6\ln\left(\frac{\mu^2}{m_H^2}\right)\right] 
  \nonumber \\ &
  +c_{h\gamma Z} \,  v_b Q_b \alpha  \, F_{h\gamma Z}\left(\frac{M_Z^2}{m_H^2},\frac{\mu^2}{m_H^2},\frac{m_b^2}{m_H^2}\right) 
  \bigg\} \, ,
 \end{align}
 where $c_{h\gamma\gamma}$ was defined in (\ref{eq:cHGG}), and 
 \begin{align}
 c_{h\gamma Z} = 2(C_{HB}-C_{HW})\hat{c}_w \hat{s}_w + C_{HWB}(\hat{c}_w^2-\hat{s}_w^2) \, , 
 \end{align}
 is the combination of Wilson coefficients entering the $h\gamma Z$ vertex in SMEFT.   
The contribution proportional to this vertex multiplies $v_b = -(\frac12 +2 Q_b \hat{s}_w^2)/(2\hat{c}_w \hat{s}_w)$, which is the vector coupling of the $Z$-boson to $b$-quarks in the SM, as well as 
 a new function $F_{h\gamma Z}$.  For arbitrary values of its arguments it is given by 
\begin{align}
\label{eq:FhAZFull}
F_{h\gamma Z} \left(z , \hat{\mu}^2, b \right) &= \frac{3}{4} \beta (8 z-5)-\beta^3 \left(\frac{39}{4}+\frac{z}{b} \right)-\frac{4}{3} \beta^2 \pi^2 \overline{z}+\frac{4}{3} \pi^2 z \overline{z} + 6 \beta \bigg( \beta^2 - \frac{2}{3} z \notag \\*  
	 &+ \left. \frac{(2 b-\beta^2) z^2}{12 b^2} \right) \ln (b)+ 2 (\beta^2
-z) \overline{z} \ln(x_z)^2 - 4\beta_z z \overline{z} \ln(x_{\beta z})  \notag \\*
	&+ \ln(x) \left( -\frac{1}{8} \left(15+7 \beta^4 + 8 z(4z-7)+ \beta^2 (2+ 8 z) \right)+ 2(z-\beta^2) \overline{z} \ln(x_z) \notag \right. \\*
	&+ 4(\beta^2-z) \overline{z} \ln(1-x x_z)+ 2(\beta^2-z) \overline{z} \ln(x_{\beta z}) \bigg) \notag \\*
	&+ \ln(x_z) \left( \frac{ \beta \beta_z z \left(\beta^2(2 b+z)-2 b z \right)}{2b^2} +2(z-\beta^2) \overline{z} \ln(x_{\beta z}) \right) \notag \\*
	&+ 4 \beta z \overline{z} \ln(\overline{z})+ \frac{ \beta^3(\beta^2+2b) z^2 \ln(z)}{2 b^2}- 6 \beta^3 \ln \left(\hat{\mu}^2 \right) \notag \\*
	&+ 4( \beta^2 -z) \overline{z} \left( \text{Li}_2 \left( \frac{x}{x_z} \right)+\text{Li}_2 \left( x x_z\right) \right) \,,
\end{align}
where 
\begin{align}
 \beta = \sqrt{1-4b} \, , \quad  \beta_z = \sqrt{1-\frac{4b}{z}}\, , \quad
x = \frac{1 -\beta}{1+\beta}, \quad x_z = \frac{1- \beta_z}{1+\beta_z} \, , \quad  x_{\beta z} = \frac{\beta-\beta_z}{\beta+\beta_z} \,  , \quad \overline{z} = 1-z \, .
\end{align}
In our numerical analysis, we use the $m_b\to 0$ limit of the above result.  The function is finite in
this limit and simplifies to 
\begin{align}
& F_{h\gamma Z} \left(z, \hat{\mu}^2, 0 \right) =
-12+ 4z-\frac{4}{3} \pi^2 \bar{z}^2 +  \left(3+ 2z + 2\bar{z}^2 \ln (\bar{z})\right)\ln(z) 
+ 4 \bar{z}^2 {\rm Li}_2(z)  
	- 6\ln(\hat{\mu}^2) \,  .
\end{align}
\subsection{Large-$m_t$ corrections}
\label{sec:LMT}
The large-$m_t$ limit of the virtual corrections to the decay rate in SMEFT has been calculated in \cite{Gauld:2015lmb}.  However, those results were limited to the on-shell scheme, and used (\ref{eq:GF}) rather than
(\ref{eq:vevrep}) to eliminate $v_T$, as appropriate in the $G_F$ scheme.  In this section we remove the restriction
to the on-shell scheme, which requires the inclusion of tadpoles, and also give results where $M_W$ instead of 
$G_F$ is used as an input parameter.  

We write the SM result as
\begin{align}
\Gamma^{(4,1)}_t=  \left[\Gamma_t^{\rm O.S.}\right]^{(4,1)}-  2 \bar{c}_{m_b}\delta b_t^{(4)}\Gamma^{(4,0)}  \,,
\end{align}
and that in SMEFT as 
\begin{align}
 \Gamma^{(6,1)}_t & =  \left[\Gamma_t^{\rm O.S.}\right]^{(6,1)}
   -2  \Gamma^{(4,0)}
 \left( \bar{c}_{m_b} \left[\delta b_t^{(6)} + 
 \delta b_t^{(4)}\left( \frac{C_{bH}\hat{v}_T^2}{\sqrt{2}}\frac{\hat{v}_T}{m_b}+
 \frac{\Gamma^{(6,0)}}{\Gamma^{(4,0)}} \right)\right]+\bar{c}_{e} \delta e_t^{(6)} 
\right)  \, ,
\end{align}
where  we have $\bar{c}_{X} \equiv 1-  c_{X}$ with $X\in \{m_b,e\}$. 
The quantity $ \Gamma^{\rm O.S.}_t$ is the decay rate 
renormalized in the on-shell scheme for $m_b$ and $e$.  The SM and dimension-6 contributions are
\begin{align}
  \left[\Gamma_t^{\rm O.S.}\right]^{(4,1)}& = \Gamma^{(4,0)} \left(-6+ N_c \frac{7-10 \hat{c}_w^2}{3 \hat{s}_w^2}\right)\frac{m_t^2}{16\pi^2 \hat{v}_T^2} \, ,  \\
 \left[\Gamma_t^{\rm O.S.}\right]^{(6,1)}  & =
\Gamma^{(6,0)}  \frac{  \left[\Gamma_t^{\rm O.S.}\right]^{(4,1)}}{\Gamma^{(4,0)}}  -  \frac{1}{2}\dot{\Gamma}_t^{(6,0)}\ln\left(\frac{\mu^2}{m_t^2}\right)  
 \nonumber \\ &
+\Gamma^{(4,0)}\frac{m_t^2}{16\pi^2} \bigg\{ C_{H\Box} N_c \frac{2+4 \hat{c}_w^2}{3\hat{s}_w^2}
- C_{HD}\left(\frac{3\hat{c}_w^2}{\hat{s}_w^2}+ N_c \frac{1+2 \hat{c}_w^4}{6 \hat{s}_w^4}\right)
 \nonumber \\ &
 +C_{HWB}\frac{\hat{c}_w}{\hat{s}_w} \left(  -12+ N_c \frac{5-8 \hat{c}_w^2}{3\hat{s}_w^2}\right)
  + \frac{C_{bH}}{\sqrt{2}}\frac{\hat{v}_T}{m_b} 
   \left(  -\frac{17}{2} + 3 N_c \frac{1-2 \hat{c}_w^2}{\hat{s}_w^2}\right)
    \nonumber \\ &
   +2  C_{Hq}^{(3)}  \left(  -1+  N_c \frac{1- 2\hat{c}_w^2}{\hat{s}_w^2}\right) \bigg\} \, .
\end{align}
The $\mu$-dependence is governed by
\begin{align}
\label{eq:GamDotDef}
\dot{\Gamma}_t^{(6,0)}\equiv {\Gamma}^{(6,0)}\big |_{C_i\to \dot{C}^{t}_{i}} \, , \quad
 \dot{C}^{t}_{i} \equiv  \frac{dC_i}{d\ln\mu}\bigg |_{m_t\to \infty} \,,
  \end{align}
where the results for $\dot{C}_i^t$ can be found in  \cite{Gauld:2015lmb}.
It is convenient to split the terms from mass and electric charge renormalization  into 
tadpole and the remaining contributions as 
\begin{align}
\label{eq:DBT}
\delta b_t &= \frac{m_t^2}{16\pi^2\hat{v}_T^2}\left( \delta\hat{b}_t + \frac{m_t^2}{m_H^2} \delta \hat{b}_{t,{\rm tad}} \right) \, ,    \\
\delta e_t & = \frac{m_t^2}{16\pi^2\hat{v}_T^2} \left( \delta\hat{e}_t + \frac{m_t^2}{m_H^2} \delta \hat{e}_{t,{\rm tad}} \right) \, .
\end{align} 
The quantities $\delta\hat{b}_t$ and  $\delta\hat{e}_t$ have been calculated in 
\cite{Gauld:2015lmb}, and are given by
\begin{align}
\delta \hat{b}_{t}^{(4)}&   = 
-\frac{5}{4}-\frac{3}{2}\ln\left(\frac{\mu^2}{m_t^2} \right) \, ,\\
\delta \hat{b}_{t}^{(6)}& =   
\hat{v}_T^2
\bigg\{ \delta\hat{b}_{t}^{(4)}\left( C_{HD} \frac{\hat{c}_w^2}{2\hat{s}_w^2}+ 2 C_{HWB}\frac{\hat{c}_w}{\hat{s}_w}+ 2 C_{Hq}^{(3)}\right)
\nonumber \\
& \qquad+  \frac{m_t}{m_b}\left(C_{Htb}+ C^{(1)}_{qtqb}(1+2N_c)+ C_F C^{(8)}_{qtqb}\right)\left[1+\ln\left(\frac{\mu^2}{m_t^2}\right)\right]\bigg\} \, ,\\
\delta \hat{e}_{t}^{(4)}& =\delta \hat{e}_{t}^{(6)}=0 \,.
\end{align}
The tadpole contributions are new, and read
\begin{align}
\delta \hat{b}_{t,{\rm tad}}^{(4)}&   = 
4N_c \left[1+ \ln\left(\frac{\mu^2}{m_t^2}\right)\right] \, ,
\\
\delta \hat{b}_{t,{\rm tad}}^{(6)} &= 2 \delta \hat{b}_{t,{\rm tad}}^{(4)} 
\left[ C_{H\Box}-\frac{C_{HD}}{4}\left(1- \frac{\hat{c}_w^2}{\hat{s}_w^2} \right)
+\frac{\hat{c}_w}{\hat{s}_w} C_{HWB}  
-\frac{\hat{v}_T}{m_b} \frac{C_{bH}}{2\sqrt{2}}
\right]\hat{v}_T^2  \, ,\\
\delta \hat{e}_{t,{\rm tad}}^{(4)}&   = 0 \, ,  \\ 
\delta \hat{e}_{t,{\rm tad}}^{(6)}&   = 
8 N_c \left[ C_{HB} \hat{c}_w^2 + C_{HW} \hat{s}_w^2 - C_{HWB} \hat{c}_w \hat{s}_w\right]\hat{v}_T^2 \left[1+ \ln\left(\frac{\mu^2}{m_t^2}\right)\right]   \, .
\end{align}

\newpage
\begin{table}
\begin{center}
\small
\begin{minipage}[t]{4.45cm}
\renewcommand{\arraystretch}{1.5}
\begin{tabular}[t]{c|c}
\multicolumn{2}{c}{$1:X^3$} \\
\hline
$Q_G$                & $f^{ABC} G_\mu^{A\nu} G_\nu^{B\rho} G_\rho^{C\mu} $ \\
$Q_{\widetilde G}$          & $f^{ABC} \widetilde G_\mu^{A\nu} G_\nu^{B\rho} G_\rho^{C\mu} $ \\
$Q_W$                & $\epsilon^{IJK} W_\mu^{I\nu} W_\nu^{J\rho} W_\rho^{K\mu}$ \\ 
$Q_{\widetilde W}$          & $\epsilon^{IJK} \widetilde W_\mu^{I\nu} W_\nu^{J\rho} W_\rho^{K\mu}$ \\
\end{tabular}
\end{minipage}
\begin{minipage}[t]{2.7cm}
\renewcommand{\arraystretch}{1.5}
\begin{tabular}[t]{c|c}
\multicolumn{2}{c}{$2:H^6$} \\
\hline
$Q_H$       & $(H^\dag H)^3$ 
\end{tabular}
\end{minipage}
\begin{minipage}[t]{5.1cm}
\renewcommand{\arraystretch}{1.5}
\begin{tabular}[t]{c|c}
\multicolumn{2}{c}{$3:H^4 D^2$} \\
\hline
$Q_{H\Box}$ & $(H^\dag H)\Box(H^\dag H)$ \\
$Q_{H D}$   & $\ \left(H^\dag D_\mu H\right)^* \left(H^\dag D_\mu H\right)$ 
\end{tabular}
\end{minipage}
\begin{minipage}[t]{2.7cm}

\renewcommand{\arraystretch}{1.5}
\begin{tabular}[t]{c|c}
\multicolumn{2}{c}{$5: \psi^2H^3 + \hbox{h.c.}$} \\
\hline
$Q_{eH}$           & $(H^\dag H)(\bar l_p e_r H)$ \\
$Q_{uH}$          & $(H^\dag H)(\bar q_p u_r \widetilde H )$ \\
$Q_{dH}$           & $(H^\dag H)(\bar q_p d_r H)$\\
\end{tabular}
\end{minipage}

\vspace{0.25cm}

\begin{minipage}[t]{4.7cm}
\renewcommand{\arraystretch}{1.5}
\begin{tabular}[t]{c|c}
\multicolumn{2}{c}{$4:X^2H^2$} \\
\hline
$Q_{H G}$     & $H^\dag H\, G^A_{\mu\nu} G^{A\mu\nu}$ \\
$Q_{H\widetilde G}$         & $H^\dag H\, \widetilde G^A_{\mu\nu} G^{A\mu\nu}$ \\
$Q_{H W}$     & $H^\dag H\, W^I_{\mu\nu} W^{I\mu\nu}$ \\
$Q_{H\widetilde W}$         & $H^\dag H\, \widetilde W^I_{\mu\nu} W^{I\mu\nu}$ \\
$Q_{H B}$     & $ H^\dag H\, B_{\mu\nu} B^{\mu\nu}$ \\
$Q_{H\widetilde B}$         & $H^\dag H\, \widetilde B_{\mu\nu} B^{\mu\nu}$ \\
$Q_{H WB}$     & $ H^\dag \sigma^I H\, W^I_{\mu\nu} B^{\mu\nu}$ \\
$Q_{H\widetilde W B}$         & $H^\dag \sigma^I H\, \widetilde W^I_{\mu\nu} B^{\mu\nu}$ 
\end{tabular}
\end{minipage}
\begin{minipage}[t]{5.2cm}
\renewcommand{\arraystretch}{1.5}
\begin{tabular}[t]{c|c}
\multicolumn{2}{c}{$6:\psi^2 XH+\hbox{h.c.}$} \\
\hline
$Q_{eW}$      & $(\bar l_p \sigma^{\mu\nu} e_r) \sigma^I H W_{\mu\nu}^I$ \\
$Q_{eB}$        & $(\bar l_p \sigma^{\mu\nu} e_r) H B_{\mu\nu}$ \\
$Q_{uG}$        & $(\bar q_p \sigma^{\mu\nu} T^A u_r) \widetilde H \, G_{\mu\nu}^A$ \\
$Q_{uW}$        & $(\bar q_p \sigma^{\mu\nu} u_r) \sigma^I \widetilde H \, W_{\mu\nu}^I$ \\
$Q_{uB}$        & $(\bar q_p \sigma^{\mu\nu} u_r) \widetilde H \, B_{\mu\nu}$ \\
$Q_{dG}$        & $(\bar q_p \sigma^{\mu\nu} T^A d_r) H\, G_{\mu\nu}^A$ \\
$Q_{dW}$         & $(\bar q_p \sigma^{\mu\nu} d_r) \sigma^I H\, W_{\mu\nu}^I$ \\
$Q_{dB}$        & $(\bar q_p \sigma^{\mu\nu} d_r) H\, B_{\mu\nu}$ 
\end{tabular}
\end{minipage}
\begin{minipage}[t]{5.4cm}
\renewcommand{\arraystretch}{1.5}
\begin{tabular}[t]{c|c}
\multicolumn{2}{c}{$7:\psi^2H^2 D$} \\
\hline
$Q_{H l}^{(1)}$      & $(H^\dag i\overleftrightarrow{D}_\mu H)(\bar l_p \gamma^\mu l_r)$\\
$Q_{H l}^{(3)}$      & $(H^\dag i\overleftrightarrow{D}^I_\mu H)(\bar l_p \sigma^I \gamma^\mu l_r)$\\
$Q_{H e}$            & $(H^\dag i\overleftrightarrow{D}_\mu H)(\bar e_p \gamma^\mu e_r)$\\
$Q_{H q}^{(1)}$      & $(H^\dag i\overleftrightarrow{D}_\mu H)(\bar q_p \gamma^\mu q_r)$\\
$Q_{H q}^{(3)}$      & $(H^\dag i\overleftrightarrow{D}^I_\mu H)(\bar q_p \sigma^I \gamma^\mu q_r)$\\
$Q_{H u}$            & $(H^\dag i\overleftrightarrow{D}_\mu H)(\bar u_p \gamma^\mu u_r)$\\
$Q_{H d}$            & $(H^\dag i\overleftrightarrow{D}_\mu H)(\bar d_p \gamma^\mu d_r)$\\
$Q_{H u d}$ + h.c.   & $i(\widetilde H ^\dag D_\mu H)(\bar u_p \gamma^\mu d_r)$\\
\end{tabular}
\end{minipage}

\vspace{0.25cm}

\begin{minipage}[t]{4.75cm}
\renewcommand{\arraystretch}{1.5}
\begin{tabular}[t]{c|c}
\multicolumn{2}{c}{$8:(\bar LL)(\bar LL)$} \\
\hline
$Q_{ll}$        & $(\bar l_p \gamma_\mu l_r)(\bar l_s \gamma^\mu l_t)$ \\
$Q_{qq}^{(1)}$  & $(\bar q_p \gamma_\mu q_r)(\bar q_s \gamma^\mu q_t)$ \\
$Q_{qq}^{(3)}$  & $(\bar q_p \gamma_\mu \sigma^I q_r)(\bar q_s \gamma^\mu \sigma^I q_t)$ \\
$Q_{lq}^{(1)}$                & $(\bar l_p \gamma_\mu l_r)(\bar q_s \gamma^\mu q_t)$ \\
$Q_{lq}^{(3)}$                & $(\bar l_p \gamma_\mu \sigma^I l_r)(\bar q_s \gamma^\mu \sigma^I q_t)$ 
\end{tabular}
\end{minipage}
\begin{minipage}[t]{5.25cm}
\renewcommand{\arraystretch}{1.5}
\begin{tabular}[t]{c|c}
\multicolumn{2}{c}{$8:(\bar RR)(\bar RR)$} \\
\hline
$Q_{ee}$               & $(\bar e_p \gamma_\mu e_r)(\bar e_s \gamma^\mu e_t)$ \\
$Q_{uu}$        & $(\bar u_p \gamma_\mu u_r)(\bar u_s \gamma^\mu u_t)$ \\
$Q_{dd}$        & $(\bar d_p \gamma_\mu d_r)(\bar d_s \gamma^\mu d_t)$ \\
$Q_{eu}$                      & $(\bar e_p \gamma_\mu e_r)(\bar u_s \gamma^\mu u_t)$ \\
$Q_{ed}$                      & $(\bar e_p \gamma_\mu e_r)(\bar d_s\gamma^\mu d_t)$ \\
$Q_{ud}^{(1)}$                & $(\bar u_p \gamma_\mu u_r)(\bar d_s \gamma^\mu d_t)$ \\
$Q_{ud}^{(8)}$                & $(\bar u_p \gamma_\mu T^A u_r)(\bar d_s \gamma^\mu T^A d_t)$ \\
\end{tabular}
\end{minipage}
\begin{minipage}[t]{4.75cm}
\renewcommand{\arraystretch}{1.5}
\begin{tabular}[t]{c|c}
\multicolumn{2}{c}{$8:(\bar LL)(\bar RR)$} \\
\hline
$Q_{le}$               & $(\bar l_p \gamma_\mu l_r)(\bar e_s \gamma^\mu e_t)$ \\
$Q_{lu}$               & $(\bar l_p \gamma_\mu l_r)(\bar u_s \gamma^\mu u_t)$ \\
$Q_{ld}$               & $(\bar l_p \gamma_\mu l_r)(\bar d_s \gamma^\mu d_t)$ \\
$Q_{qe}$               & $(\bar q_p \gamma_\mu q_r)(\bar e_s \gamma^\mu e_t)$ \\
$Q_{qu}^{(1)}$         & $(\bar q_p \gamma_\mu q_r)(\bar u_s \gamma^\mu u_t)$ \\ 
$Q_{qu}^{(8)}$         & $(\bar q_p \gamma_\mu T^A q_r)(\bar u_s \gamma^\mu T^A u_t)$ \\ 
$Q_{qd}^{(1)}$ & $(\bar q_p \gamma_\mu q_r)(\bar d_s \gamma^\mu d_t)$ \\
$Q_{qd}^{(8)}$ & $(\bar q_p \gamma_\mu T^A q_r)(\bar d_s \gamma^\mu T^A d_t)$\\
\end{tabular}
\end{minipage}

\vspace{0.25cm}

\begin{minipage}[t]{3.75cm}
\renewcommand{\arraystretch}{1.5}
\begin{tabular}[t]{c|c}
\multicolumn{2}{c}{$8:(\bar LR)(\bar RL)+\hbox{h.c.}$} \\
\hline
$Q_{ledq}$ & $(\bar l_p^j e_r)(\bar d_s q_{tj})$ 
\end{tabular}
\end{minipage}
\begin{minipage}[t]{5.5cm}
\renewcommand{\arraystretch}{1.5}
\begin{tabular}[t]{c|c}
\multicolumn{2}{c}{$8:(\bar LR)(\bar L R)+\hbox{h.c.}$} \\
\hline
$Q_{quqd}^{(1)}$ & $(\bar q_p^j u_r) \epsilon_{jk} (\bar q_s^k d_t)$ \\
$Q_{quqd}^{(8)}$ & $(\bar q_p^j T^A u_r) \epsilon_{jk} (\bar q_s^k T^A d_t)$ \\
$Q_{lequ}^{(1)}$ & $(\bar l_p^j e_r) \epsilon_{jk} (\bar q_s^k u_t)$ \\
$Q_{lequ}^{(3)}$ & $(\bar l_p^j \sigma_{\mu\nu} e_r) \epsilon_{jk} (\bar q_s^k \sigma^{\mu\nu} u_t)$
\end{tabular}
\end{minipage}
\end{center}
\caption{\label{op59}
The 59 independent baryon number conserving dimension-6 operators built from Standard Model fields, in 
the notation of \cite{Jenkins:2013zja}.  The subscripts $p,r,s,t$ are flavour indices, and $\sigma^I$ are Pauli
matrices.}
\end{table}

\end{document}